\documentclass[preprint]{elsarticle}

\usepackage[utf8]{inputenc}
\usepackage{todonotes}
\usepackage{url}
\usepackage{amsmath}
\usepackage{amssymb}
\usepackage{xspace}
\usepackage{siunitx}
\usepackage{pdflscape}
\usepackage{chemformula}
\usepackage{multirow}
\usepackage{subcaption}
\usepackage[section]{placeins}
\usepackage{lineno}
\usepackage[colorlinks=true,allcolors=blue]{hyperref}
\usepackage[capitalize]{cleveref}
\usepackage{stmaryrd}
\usepackage{booktabs}
\usepackage{listings}
\definecolor{comment}{rgb}{0.0,0.6,0.0}

\lstdefinestyle{cppstyle}{language=C++, basicstyle=\ttfamily,
  extendedchars=true, escapeinside={/*@}{@*/},
  breaklines=true, breakatwhitespace=true,
  numbers=left, numberstyle=\tiny,
  xleftmargin=1pt,
  keywordstyle=\color{blue}\bfseries,
  stringstyle=\color{red}\ttfamily,
  commentstyle=\color{comment}\ttfamily,
  morecomment=[l][\color{magenta}]{\#},
  captionpos=b,
  abovecaptionskip=10pt
}
\lstnewenvironment{c++}[1][]{\lstset{style=cppstyle}}{}
\newcommand{\cpp}{\lstinline}
\crefname{lstlisting}{Code Example}{Code Example}
\Crefname{lstlisting}{Code Example}{Code Example}

\usepackage{tikz}
\usetikzlibrary{shapes.geometric, arrows}

\definecolor{mygreen}{RGB}{136,170,0}
\definecolor{myblue}{RGB}{85,85,255}
\tikzstyle{greencircle} =    [circle, minimum width=1cm, minimum height=1cm, text centered, draw=white, text=white, align=center, fill=mygreen]
\tikzstyle{greenrectangle} = [rectangle, minimum width=1cm, minimum height=1cm, text centered, draw=white, text=white, align=center, fill=mygreen]
\tikzstyle{bluecircle} =     [circle, minimum width=1cm, minimum height=1cm, text centered, draw=white, text=white, align=center, fill=myblue]
\tikzstyle{bluerectangle} =  [rectangle, minimum width=1cm, minimum height=1cm, text centered, draw=white, align=center, fill=myblue]
\tikzstyle{arrow} = [thick,->,>=stealth]
\tikzstyle{doublearrow} = [thick,<->,>=stealth]

\numberwithin{equation}{section}

\graphicspath{{./img/}}

\newcommand{\dune}{\textsc{Dune}\xspace}
\newcommand{\dumux}{DuMu\textsuperscript{x}\xspace}
\newcommand{\dunemodule}[1]{\texttt{#1}\xspace}

\newcommand{\someStorage}{m}
\newcommand{\someFlux}{\boldsymbol{\psi}}
\newcommand{\someSource}{q}
\newcommand{\domain}{\Omega}
\newcommand{\domainBoundary}{\partial\Omega}
\newcommand{\neumannIdx}{\mathrm{N}}
\newcommand{\dirichletIdx}{\mathrm{D}}
\newcommand{\R}{\mathbb{R}}

\newcommand{\element}{E}
\newcommand{\mesh}{\mathcal{M}}
\newcommand{\cvSet}{\mathcal{T}}
\newcommand{\faceSet}{\mathcal{E}}
\newcommand{\cv}{K}
\newcommand{\cvtwo}{L}
\newcommand{\scv}{\kappa}
\newcommand{\scvf}{\sigma}
\newcommand{\n}{\mathbf{n}}
\newcommand{\discFlux}{F}

\newcommand{\porosity}{\phi}
\newcommand{\permeability}{\mathbf{K}}
\newcommand{\pressure}{p}
\newcommand{\saturation}{S}
\newcommand{\darcyVelocity}{\mathbf{v}}
\newcommand{\fluidPhaseIdx}{\beta}
\newcommand{\relPerm}[1]{k_{\text{r} #1}}
\newcommand{\dynVisc}{\mu}
\newcommand{\density}{\rho}
\newcommand{\gravity}{\mathbf{g}}

\newcommand{\anisotropy}{\Xi}
\newcommand{\permangle}{\Phi}

\newcommand{\aperture}{a}

\newcommand{\water}{ {\mathrm{H}_2\mathrm{O}} }
\newcommand{\nitrogen}{ {\mathrm{N}_2} }

\renewcommand{\div}{\nabla\cdot\!}
\newcommand{\scal}{\cdot}
\newcommand{\grad}{\nabla\!}
\newcommand{\gradt}[1]{\left(\nabla\!#1\right)^T}

\newcommand{\meas}[1]{\lvert{#1}\rvert}  % the measure of something

\begin{document}

\title{\dumux 3 -- an open-source simulator for solving flow and transport problems in porous media with a focus on model coupling}
\author[1]{Timo Koch\corref{cor1}}
\ead{timo.koch@iws.uni-stuttgart.de}
\author[1]{Dennis Gläser}
\author[1]{Kilian Weishaupt}

\author[1]{Sina Ackermann}
\author[1]{Martin Beck}
\author[1]{Beatrix Becker}
\author[1]{Samuel Burbulla}
\author[1]{Holger Class}
\author[1]{Edward Coltman}
\author[1]{Simon Emmert}
\author[1]{Thomas Fetzer}
\author[1]{Christoph Grüninger}
\author[1]{Katharina Heck}
\author[1]{Johannes Hommel}
\author[1]{Theresa Kurz}
\author[1]{Melanie Lipp}
\author[1]{Farid Mohammadi}
\author[3]{Samuel Scherrer}
\author[1]{Martin Schneider}
\author[1]{Gabriele Seitz}
\author[4]{Leopold Stadler}
\author[4]{Martin Utz}
\author[1]{Felix Weinhardt}
\author[1]{Bernd Flemisch}

\cortext[cor1]{Corresponding author}
\address[1]{Department of Hydromechanics and Modelling of Hydrosystems, Institute for Modelling Hydraulic and Environmental Systems, University of Stuttgart, 70569 Stuttgart, Germany}
\address[2]{Chair of Applied Mathematics, Institute of Applied Analysis and Numerical Simulation, University of Stuttgart, 70569 Stuttgart, Germany}
\address[3]{VEGAS - Research Facility for Subsurface Remediation, Institute for Modelling Hydraulic and Environmental Systems, University of Stuttgart, 70569 Stuttgart, Germany}
\address[4]{BAW - Federal Waterways Engineering and Research Institute, 76187 Karlsruhe, Germany}

\begin{abstract}
We present version 3 of the open-source simulator for flow and transport processes in porous media \dumux.
\dumux is based on the modular C++ framework \dune (Distributed and Unified Numerics Environment)
and is developed as a research code with a focus on modularity and reusability.
We describe recent efforts in improving the transparency and efficiency of the development process and community-building,
as well as efforts towards quality assurance and reproducible research.
In addition to a major redesign of many simulation components in order to facilitate setting up
complex simulations in \dumux, version 3 introduces a more consistent abstraction of finite volume schemes.
Finally, the new framework for multi-domain simulations is described, and three numerical examples demonstrate its flexibility.
\end{abstract}


\maketitle

\section{Introduction}
\label{sec:intro}

\dumux, \dune for multi-\{phase, component, scale, physics, domain \ldots \} flow and transport in porous media, is a free and open-source simulator for flow and transport processes in porous media \cite{FlemischEtAl:2011:DDM} (\href{http://dumux.org}{\texttt{dumux.org}}). Its main intention is to provide a sustainable, consistent and modular framework for the implementation and application of porous media model concepts and constitutive relations. It has been successfully applied to greenhouse gas and CO\textsubscript{2} storage~\citep{Nordbotten:2012:UPS,ahusborde2015numerical,hagemann2016hydrogenization,Walter_al_IJGGC2012},
radioactive waste disposal \citep{ahusborde2015three},
environmental remediation problems \citep{weishaupt2016numerical},
transport of therapeutic agents through biological tissue \cite{erbertseder2012coupled,Koch2018a,Vidotto2018,stoverud2012modeling},
fractured porous media \citep{Schwenck:2015:DRF,stadler2012modeling,tecklenburg2016multi,glaser2017discrete},
and subsurface-atmosphere coupling \citep{MosthafEtAl2011,fetzer2016effect}.
For a more complete list of publications that have been achieved with the help of \dumux, see \href{https://dumux.org/publications}{\url{dumux.org/publications}}.

\dumux is based on the Distributed Unified Numerics Environment (\dune)~\citep{dunegrid2,dunegrid1,dune-web-page}, an open-source scientific numerical software framework for solving partial differential equations, and is thus part of a larger community that goes beyond the simulation of fluid-mechanical processes in porous media. \dune and \dumux are written in C++, using modern C++ programming techniques and C++ template meta programming for efficiency and generic interfaces. The \dune core modules provide, among other things, multiple grid managers implementing a versatile common grid interface, linear algebra abstraction and an iterative solver back-end~\citep{duneistl}, as well as abstractions facilitating parallel computing.
\dumux is designed and developed as a \dune module depending on the \dune core modules and optionally interacts with a number of other \dune-based extension modules.
The key features of \dumux are its comprehensive library of multi-phase and multi-component flow and transport models, the flexible and modular fluid and
material framework for constitutive relations~\citep{FlemischEtAl:2011:DDM}, abstractions for finite volume discretization schemes (see~\cref{sec:fv}),
and its focus on model coupling (see~\cref{sec:multidomain}).

There are many other open-source projects focusing on Darcy-scale porous-medium flow and transport processes, such as
MODFLOW, \href{https://water.usgs.gov/ogw/modflow/}{\url{
water.usgs.gov/ogw/modflow/}} \citep{McDonald:2003:HMO},
MRST, \href{https://www.sintef.no/projectweb/mrst/}{\url{
sintef.no/projectweb/mrst/}} \citep{lie_2019},
OpenGeoSys, \href{http://www.opengeosys.org}{\url{opengeosys.org}} \citep{Kolditz2012OpenGeoSys},
OPM, \href{https://opm-project.org/}{\url{opm-project.org}} \citep{Baxendale:2018:OPM},
ParFlow, \href{https://parflow.org/}{\url{parflow.org}} \citep{maxwell2015high},
PFloTran, \href{http://www.pflotran.org/}{\url{pflotran.org}} \citep{lichtner2015pflotran},
or
PorePy, \href{https://github.com/pmgbergen/porepy}{\url{github.com/pmgbergen/porepy}} \citep{Keilegavlen2017}.
Furthermore, there are open-source numerical software frameworks with a broader focus such as
deal.II, \href{http://www.dealii.org/}{\url{dealii.org}} \citep{bangerth2007deal},
\dune, \href{https://www.dune-project.org/}{\url{dune-project.org}} \citep{dunegrid1,dunegrid2},
Feel++, \href{https://www.feelpp.org}{\url{feelpp.org}} \citep{prud2012feel},
FEniCS, \href{https://fenicsproject.org/}{\url{fenicsproject.org}} \citep{logg2012automated},
MOOSE, \href{http://www.mooseframework.org/}{\url{mooseframework.org}} \citep{Gaston2009},
or
OpenCMISS, \href{https://opencmiss.org/}{\url{opencmiss.org}} \citep{bradley2011opencmiss}.

This paper can be seen as a continuation of \cite{FlemischEtAl:2011:DDM} which describes the code and the project at the beginning of the 2.X release series.
In the remainder of this introductory section, we provide a brief chronology of the development of \dumux
and some information on activities and measures that go beyond the development of the mainline code base.
\Cref{sec:design} introduces the general structure of the code and its design principles.
\cref{sec:fv} goes into more details by explaining abstractions and software concepts for general finite volume schemes encountered in \dumux.
\Cref{sec:multidomain} is devoted to simulations coupling two or more computational domains and models.
In \cref{sec:features}, we briefly address new features in the new version \dumux 3.
Numerical examples that particularly highlight the strengths of the new multi-domain framework are presented in \cref{sec:gallery}.
Finally, \cref{sec:future} discusses current limitations and perspectives.

\subsection{History}
The following section briefly outlines the \dumux project history and its development. More details can be found in \citep{Bilke:2019:DOS}.

\paragraph{Initial development and first release}
The development of \dumux started in January 2007 at the Department of Hydromechanics and Modelling of Hydrosystems (LH2) at the University of Stuttgart.
After an evaluation of the PDE software frameworks available at that time, the decision fell to build the code on top of the C++ toolbox \dune~\citep{dunegrid1,dunegrid2}.
In March 2007, a Subversion repository was set up for \dumux to host and control the code development.
From that point on, every new doctoral student and postdoc at the LH2 performed his/her modeling tasks by using and enhancing the program code in that repository.
Up until today, most developers of \dumux are associated to the LH2 working group.
This naturally results in a continuous major goal of the project that is to provide a tool enabling all developers to perform their research and possibly also teaching tasks.
In July 2009, \dumux 1.0 was released under the GNU General Public License Version 2 (or later)~\citep{GPL}.

\paragraph{The 2.X release series}
\dumux 1.0 consisted of a subset of the code stored in the Subversion repository because not everything in there was adequate for public release. Since selecting a subset of a private code base for public release proved to be rather impractical, the repository was split into a stable part \texttt{dumux} and a development part \texttt{dumux-devel} that was dependent on the stable part. Public read access to the repository of the stable part was granted.
Since the code design still exhibited several shortcomings concerning dependencies, generality and modularity, a major refactoring of the code base was performed in the following 1.5 years. As a result, \dumux 2.0 was released in February 2011 which also lead to the main reference \cite{FlemischEtAl:2011:DDM}. After this release, interfaces were kept stable for at least one release cycle in order to achieve more sustainability and security for the growing number of users and developers.

Following the tendency from centralized to distributed version control, the \dumux Subversion repository was converted into Git repositories in September 2015, following the transition of the \dune framework to Git.
GitLab was employed as a version control management system, and the Git repositories are publicly accessible at \href{https://git.iws.uni-stuttgart.de}{\texttt{git.iws.uni-stuttgart.de}}.
The release process has been streamlined and since release 2.4 in October 2013, \dumux has been released semiannually in spring and autumn every year, the last release of the 2.X series being 2.12 in late 2017.
Since 2.7, every release tarball is uploaded to Zenodo, thereby receiving a DOI, as for example~\citep{dumux3zenodo}.

\paragraph{Transition to \dumux 3}
During the 2.X release series, several new features were added to the code base.
Many of these additions fell in line with the main intention of \dumux to be a framework for the
implementation of porous-media model concepts and constitutive relations by actually providing
such implementations. However, some more central additions had to be rather forced into
the code base and proved to be inefficient, inconsistent with the original design ideas or increasingly difficult to maintain.
This lead to a new major release cycle that was initiated by branching off the main development line in November 2016.
Due to the large amount of changes and their extent into all parts of the code base,
the requirement of backward compatibility was dropped.
Right after the release of \dumux 2.12 in December 2017, the development branch
was integrated back into the main line and an alpha release was published.
It took another year before \dumux 3.0 finally was released in December 2018.
Starting with the release of \dumux 3.1 scheduled for October 2019, the aforementioned semiannual release cycle is resumed.

\subsection{\dumux as a framework}
In the following, we address aspects going beyond the mainline code features and development.
After discussing quality assurance and reproducibility, we introduce the two modules \dunemodule{dumux-course} and \dunemodule{dumux-lecture}.
This is followed by a brief sketch of the Open Porous Media (OPM) initiative and looking at community building.
Again, more details are presented in \cite{Bilke:2019:DOS}.

\paragraph{Quality assurance and reproducibility}
To assure the quality of the developed software, \dumux is currently accompanied by about 400 unit and system tests
automatically built and run by a BuildBot CI server at \href{https://git.iws.uni-stuttgart.de/buildbot/}{\url{git.iws.uni-stuttgart.de/buildbot}}
after each commit to the master branch.
We regularly assess that the tests cover a large number of lines of code and work towards increasing
the coverage\footnote{A detailed weekly coverage report created with gcov, gcovr and GitLab CI is publicly
available at \href{https://pages.iws.uni-stuttgart.de/dumux-repositories/dumux-coverage/}{\url{pages.iws.uni-stuttgart.de/dumux-repositories/dumux-coverage/}}}.
One guideline for developers is that every newly added feature has to be accompanied
by a corresponding test and a comprehensive documentation.
In order to achieve reproducibility of the computational results, the project \dumux-Pub
has been initiated in 2015. It provides a set of tools for the researcher to extract
the code that has been used in a publication into a separate \dune module, together
with an install script to download and compile the code, including all necessary dependencies.
Since 2015, every journal publication at the LH2 as well as every bachelor, master and doctoral
thesis has to be accompanied by such a module. All resulting modules are published at
\href{https://git.iws.uni-stuttgart.de/dumux-pub}{\url{git.iws.uni-stuttgart.de/dumux-pub}}.
First efforts have been undertaken to provide also a complete runtime environment in form of
a Docker container, see for example \href{https://git.iws.uni-stuttgart.de/dumux-pub/Koch2017a}{\url{git.iws.uni-stuttgart.de/dumux-pub/Koch2017a}}.
These efforts are continued as part of the ongoing DFG project ``Sustainable infrastructure for the improved usability and archivability of research software on the example of the porous-media-simulator \dumux'' (SusI), which also aims to provide browser frontends to operate corresponding \dumux Docker containers.

\paragraph{\dumux course}
The \dunemodule{dumux-course} module has been created for a \dumux course offered 2018 in Stuttgart. It is continuously improved and enhanced along with the mainline development.
While the module is used in conventional \dumux courses with participants attending personally,
it is also designed to be used independently by everyone who would like to learn how to work with \dumux.
All course exercises are documented and contain task descriptions and solutions. Their contents range from a very basic first building and running experience to fairly advanced applications based on coupled model equations. The module also contains the slides from the course providing broader context and background information.

\paragraph{\dumux lecture}
The Department of Hydromechanics and Modelling of Hydrosystems at the University of Stuttgart (LH2) offers Master level courses for different study programs (Environmental and Civil Engineering, Water Resources Engineering and Management, Simulation Technology, or Computational Mechanics of Materials and Structures), where computer exercises using \dumux are an essential part. The module \dunemodule{dumux-lecture} contains all the example applications employed in these lectures, with most of the applications belonging to the lecture ``Multiphase Modeling''. Each application has its own folder containing problem setup and spatial parameter specifications, as well as an input file where typically those runtime parameters can be specified that are of educational value for the problem. Each example is accompanied by explanations and tasks.

\paragraph{The Open Porous Media (OPM) initiative}
In 2009, the OPM initiative was born with the principal objective
``to develop a simulation suite that is capable of modeling industrially and scientifically relevant flow and transport processes in porous media and bridge the gap between the different application areas of porous media modeling'' \citep{Lie:2009:OPM} (\href{https://opm-project.org}{\url{opm-project.org}}).
Currently, the main focus of OPM is on reservoir engineering with a particular emphasis on being able to compete with proprietary industry-standard tools.
As such, the black-oil simulator \texttt{Flow} is a main product of OPM. \texttt{Flow} is built upon the OPM modules \texttt{opm-material} and \texttt{ewoms}, both of which originated from a fork of \dumux 2.2 in March 2012.
In turn, \dumux has a suggested dependency on the OPM module \texttt{opm-grid} which enables \dumux users to use corner-point grids, the de facto standard in the petroleum industry.

\paragraph{Community building}
By going open source, external users of \dumux have been welcome ever since the initial release in 2009.
In order to get in contact with the users, a first \dumux user meeting was held in Stuttgart in June 2015. 26 participants were counted, 10 of them were external users. For attracting new users, a first \dumux course was given in October 2017, followed by a second one in July 2018, and a short course at the InterPore conference 2019.
In May 2019, the kick-off workshop for the aforementioned project SusI took place in Stuttgart, where the project investigators and 12 external users focused on identifying current shortcomings and possible remedies associated with the usability of \dumux.
With GitLab, all technical possibilities are available for users to upload contributions to the code base in form of merge requests as well as for developers to review, discuss, improve and possibly integrate them. However, contributions from outside the LH2 are very scarce until now. Current notable exceptions are members from the neighboring working group VEGAS and the Institute of Applied Analysis and Numerical Simulation (IANS) at the University of Stuttgart as well as researchers from the Federal Waterways Engineering and Research Institute (BAW) in Karlsruhe.
%With the apparent increase in interest and actual usage over the last years, we hope to increase as well the number of external contributors and, ultimately, developers in the future.

\section{Structure and design principles}
\label{sec:design}

\dumux is designed as a research code framework. Foremost, this means that it is designed with an emphasis on modularity.
\dumux user code is usually developed in a separate \dune module that lists \dumux as its dependency.
To maintain flexibility for a user, \dumux follows the principle
that all components of a simulation should be easily replaceable with a new implementation,
without modifying code in the \dumux module itself. For example, it is
simple to change from one of the implemented capillary-pressure relationships to a custom function,
to modify the numerical flux computation, and to change from
a cell-centered to a vertex-centered finite volume discretization.
\dumux profits from the modular design of \dune. For example, an unstructured grid implementation can be
exchanged for an efficient structured grid implementation by changing a single line in the user code.

To maintain computational efficiency, much of the modularity is realized through the use of C++ templates and generic
programming techniques such as traits and policies. Moreover, the developers of \dumux try to follow well-known object-oriented design
principles and the separation of algorithms and data structures, as for example in the design of the
\dumux material framework~\citep{Lauser2012}. \dumux 3, which requires a C++14-compatible compiler, increasingly makes
use of modern C++ features such as lambdas, type deduction, or smart pointers,
to the benefit of usability, modularity, and efficiency.

In the \dumux environment, the term \textit{model} is used to describe
a system of coupled partial differential equations (PDEs) including constitutive equations needed for closure.
Many models, in particular PDEs describing general non-isothermal multi-component multi-phase flow processes in porous media,
are already implemented in \dumux. Furthermore, a user can choose from a variety of constitutive laws to describe closure relations,
and multiple fluid and solid systems and components. More importantly, the code design also facilitates using
custom implementations.

The main components of a \dumux simulation are represented in the code by corresponding C++ classes.
When using one of the implemented models out-of-the-box, a user usually implements at least two
such classes in addition to the program's \cpp{main} function. The \cpp{Problem} class defines
boundary conditions, initial conditions (if necessary), and volumetric source terms (if any),
by implementing a defined class interface.
The \cpp{SpatialParams} class defines the spatial distribution of material parameters, such as porosity, permeability, or for example,
parameters for the van Genuchten water retention model (if applicable). The \cpp{Problem} class
stores a pointer to an instance of the \cpp{SpatialParams} class. Moreover, each simulation
currently creates at least one tag (a C++ struct) which is used to specify several compile time options
(\textit{properties}) of the simulation by means of partial template specialization. \Cref{code:properties} shows how to
set properties for a newly defined tag. Properties can be extracted as traits of such a tag in other parts of the code.
Depending on the chosen model, more of such properties have to be defined and examples
are provided for all existing models in \dumux. User defined models can provide their own properties.

\begin{lstlisting}[style=cppstyle,basicstyle=\ttfamily\scriptsize,label={code:properties},
                   caption={An example of a property setting for \dumux 3. The simulation is configured to use a one-phase porous medium model (\texttt{OneP}),
                            a cell-centered two-point-flux-approximation discretization scheme (\texttt{CCTpfaModel}), and liquid water as a fluid.
                            The \texttt{UserProblem} and \texttt{UserSpatialParams} types are implementation- and user-defined. Header includes are omitted for brevity.
                            The types attached to the \texttt{Example} tag, can be extracted as traits, see \texttt{GetPropType} in~\cref{code:main}.}]
namespace Dumux {
namespace Properties {

// create a new type tag
namespace TTag {
struct Example { using InheritsFrom = std::tuple<OneP, CCTpfaModel>; };
} // end namespace TTag

// set the Dune grid type
template<class Tag> struct Grid<Tag, TTag::Example>
{ using type = Dune::UGGrid<3>; };

// set the problem type
template<class Tag> struct Problem<Tag, TTag::Example>
{ using type = UserProblem; };

// set the spatial parameters type
template<class Tag>
struct SpatialParams<Tag, TTag::Example>
{ using type = UserSpatialParams; };

// set the fluid system type
template<class Tag> struct FluidSystem<Tag, TTag::Example>
{ using type = FluidSystems::OnePLiquid<double, Components::H2O<double> >; };

} // end namespace Properties
} // end namespace Dumux
\end{lstlisting}

A rather significant improvement in \dumux 3 concerns the way the program's \cpp{main} function is written.
In \dumux 2, the \cpp{main} function called a predefined \cpp{start} function that determined the program flow.
Modifications to the program flow were made possible by numerous hook functions.
For example, the \cpp{Problem} base class implemented a hook function
\cpp{postTimeStep} that could be overloaded by the user's \cpp{Problem} implementation, injecting
the notion of time and program flow into the \cpp{Problem} class; a hook function \cpp{addVtkOutputFields}
injected a notion of file I/O. This design concept
introduced many dependencies between classes and (as shown here for the \cpp{Problem} class)
lead to the violation of the well-known single responsibility principle, hindering modular design.
Furthermore, due to the concept of hook functions,
it was very difficult to write simulations outside the predefined program flow in \cpp{start}.
For example, all simulations were considered transient and non-linear. Consequently, solving stationary equations was often
realized by solving a single time step and making sure that the assembled storage term is zero. Solving
linear equations was realized by executing a single iteration of a Newton method.
On the positive side, a user had the possibility to realize most changes within a single header file.

In contrast, the program's \cpp{main} function in \dumux 3 contains all of the main steps of the simulation,
see~\cref{code:main}. Therefore, to solve a stationary PDE, the \cpp{TimeLoop} instance (l.~41)
and the time loop (l.~53ff) can simply be removed from the program flow. Similarly, the \cpp{NewtonSolver} class
can be replaced by a \cpp{LinearPDESolver} to solve a PDE. Finally, the I/O logic and other modifications
to the main program flow can be directly written inside the \cpp{main} function. As a result, the main program flow becomes arguably
more transparent, which has been confirmed by feedback from the \dumux user community.
For more information on the steps of a \dumux simulation,
we refer the interested reader to the \dumux handbook~\citep{dumux-handbook}.

\begin{lstlisting}[style=cppstyle,basicstyle=\ttfamily\scriptsize,label={code:main},
                   caption={An example of a \texttt{main} function using \dumux 3 for solving a transient non-linear problem with a finite volume scheme and numeric differentiation. The simulation can be run in parallel (MPI, distributed memory). Header includes are omitted for brevity.}]
int main(int argc, char** argv)
{
  using namespace Dumux;

  // an example tag from which multiple traits
  // can be extracted (see GetPropType below)
  using Tag = Properties::TTag::Example;

  // initialize MPI (enables parallel runs)
  Dune::MPIHelper::instance(argc, argv);

  // parse command line arguments and a parameter file
  Parameters::init(argc, argv, "params.input");

  // create a Dune grid (from information in the parameter file)
  using Grid = GetPropType<Tag, Properties::Grid>;
  GridManager<Grid> gridManager; gridManager.init();

  // create the finite volume grid geometry (see Section 3)
  const auto leafGridView = gridManager.grid().leafGridView();
  using GridGeometry = GetPropType<Tag, Properties::FVGridGeometry>;
  auto fvGridGeometry = std::make_shared<GridGeometry>(leafGridView);
  fvGridGeometry->update();

  // the problem (initial conditions, boundary conditions, sources)
  using Problem = GetPropType<Tag, Properties::Problem>;
  auto problem = std::make_shared<Problem>(fvGridGeometry);

  // the solution vector, set to the initial solution
  using SolutionVector = GetPropType<Tag, Properties::SolutionVector>;
  SolutionVector x; problem->applyInitialSolution(x);
  auto xOld = x;
  // the grid variables (secondary variables on control volumes and faces)
  using GridVariables = GetPropType<Tag, Properties::GridVariables>;
  auto gridVariables = std::make_shared<GridVariables>(problem, fvGridGeometry);
  gridVariables->init(x);

  // get some time loop parameters and instantiate time loop
  const auto tEnd = getParam<double>("TimeLoop.TEnd");
  auto dt = getParam<double>("TimeLoop.DtInitial");
  auto timeLoop = std::make_shared<TimeLoop<double>>(0, dt, tEnd);

  // the assembler with time loop for transient problem & the solver
  using Assembler = FVAssembler<Tag, DiffMethod::numeric>;
  auto assembler = std::make_shared<Assembler>(problem, fvGridGeometry, gridVariables, timeLoop);
  assembler->setPreviousSolution(xOld);
  Dumux::NewtonSolver<Assembler, LinearSolver> nonLinearSolver(assembler, linearSolver);

  // initialize output module
  ...

  // time loop
  timeLoop->start(); do
  {
    // solve the non-linear system with time step control
    nonLinearSolver.solve(x, *timeLoop);

    // make the new solution the old solution
    xOld = x; gridVariables->advanceTimeStep();

    // advance the time loop to the next step
    timeLoop->advanceTimeStep();

    // report statistics of this time step
    timeLoop->reportTimeStep();

    // write output, set new time step size
    ...

  } while (!timeLoop->finished());

  return 0;
}
\end{lstlisting}

\section{Abstractions and concepts for general finite volume schemes}
\label{sec:fv}

One of the most important abstractions in \dumux is the \textit{grid geometry}.
The concept partly existed in \dumux 2 but was significantly redesigned and has been casted into
an object-oriented representation in \dumux 3. A grid geometry is a wrapper around a \textit{grid view}
on a \dune \textit{grid} instance. \dune grids are general hierarchical grids and grid views provide read-only access
to certain parts of the grid. In particular, a \textit{leaf grid view} is a view on all grid entities without
descendants in the hierarchy (which are not refined), thus covering the whole domain (in sequential simulations)
or the part of the grid assigned to a single processor (in parallel simulations), while a \textit{level grid view}
is a view on all entities of a given level of the refinement hierarchy.
The grid geometry constructs, from such a grid view, all the geometrical and topological
data necessary to evaluate the discrete equations
resulting from a given finite volume scheme. This abstraction allows us to implement many different
finite volume schemes in a unified way. In the following sections, we will present and motivate the mathematical abstractions
behind the grid geometry concept, describe their realizations in the code in form of C++ classes and provide an exemplary code snippet that illustrates how to use them.

\subsection{Finite volume discretization}
\label{sec:fvmath}

Let us consider a domain $\domain \subset \R^n$, $1 \leq n \leq 3$, with boundary $\domainBoundary$, which is further decomposed into two subsets on which Dirichlet and Neumann boundaries are specified, that is $\domainBoundary = \domainBoundary_\dirichletIdx \cup \domainBoundary_\neumannIdx$. Furthermore, let us consider a general conservation equation,

\begin{subequations}
  \label{eq:someProblem}
  \begin{align}
    \frac{\partial \someStorage}{\partial t} + \div \someFlux &= \someSource, &&\mathrm{in} \, \domain,     \label{eq:someConservation}\\
               \someFlux \cdot \n &= g_\neumannIdx, &&\mathrm{on} \, \domainBoundary_\neumannIdx,                    \label{eq:someNeumannBC}\\
                                u &= u_\dirichletIdx, &&\mathrm{on} \, \domainBoundary_\dirichletIdx,                \label{eq:someDirichletBC}\\
                           u(t=0) &= u_0 &&\mathrm{in} \, \domain,                                                   \label{eq:someIC}
  \end{align}
\end{subequations}
where $\someStorage = \someStorage(u)$ is the conserved quantity depending on the primary variable $u$,
$\someFlux = \someFlux(u)$ is a flux term and $\someSource = \someSource(u)$ a source term. All terms may non-linearly depend
on $u$. \Cref{eq:someConservation,eq:someNeumannBC,eq:someDirichletBC,eq:someIC} constitute a non-linear problem in $u$ with initial and boundary conditions.

We introduce the primary grid $\mesh$ with elements $\element\in\mesh$, such that $\domain_h = \bigcup_{\element\in\mesh}\element$ is a discrete approximation of $\domain$.
Furthermore, we introduce a tessellation $\cvSet$ of $\domain$, such that $\domain_h = \bigcup_{\cv \in \cvSet} \cv$,
where each $\cv \in \cvSet$ is a control volume with measure $\meas{\cv} > 0$.
The control volumes $\cv$ do not necessarily need to coincide with the elements $\element$. However, for cell-centered finite volume schemes, usually $\cvSet\equiv\mesh$.
Each control volume is partitioned into sub-control-volumes $\scv$, where $\scv = \cv$ is one admissible partition.
Moreover, the boundary of the control volume $\partial\cv$ consists of a finite number of faces $f \subset \partial\cv$ which are either
inner faces $f_I = \partial\cv \cap \partial\cvtwo$, where $\cvtwo\in\cvSet$ is an adjacent control volume, or boundary faces
$f_B = \partial\cv \cap \partial\domain_h$. Depending on the discretization method, it may be useful to partition the faces $f$ into sub-(control-volume)-faces $\scvf$.
We denote by $\faceSet_\cv$ the entire set of such sub-faces on $\partial \cv$, forming a disjoint partition such that $\partial \cv = \bigcup_{\scvf \in \faceSet_\cv} \scvf$. Accordingly, $\cvSet_\cv$ denotes the set of sub-control volumes embedded in a control volume $\cv$ with $\cv = \bigcup_{\scv \in \cvSet_\cv} \scv$.

Integration of \cref{eq:someConservation} over a control volume and application of the divergence theorem yields
\begin{equation}
  \int_\cv \frac{\partial \someStorage}{\partial t} \, \mathrm{d} x
  + \sum_{f \subset \partial\cv} \int_f \someFlux\cdot\n \, \mathrm{d} \Gamma = \int_\cv \someSource \, \mathrm{d} x,
  \label{eq:someConservationDiscrete1}
\end{equation}
where $\n$ is a unit outer normal vector on $\partial\cv$.
Using the concept of sub-control-volumes and sub-control-volume-faces, allows to reformulate \cref{eq:someConservationDiscrete1} as
\begin{equation}
  \sum_{\scv \in \cvSet_\cv} \int_\scv \frac{\partial \someStorage}{\partial t} \, \mathrm{d} x
  + \sum_{\scvf \in \faceSet_\cv} \int_\scvf \someFlux\cdot\n \, \mathrm{d} \Gamma = \sum_{\scv \in \cvSet_\cv} \int_\scv \someSource \, \mathrm{d} x.
  \label{eq:someConservationDiscrete2}
\end{equation}

Let us now replace the exact flux over sub-face $\scvf$ by the approximation $\discFlux_{\cv, \scvf}\approx\int_\scvf \someFlux\cdot\n\,\mathrm{d}\Gamma$
and approximate the volume integrals in \cref{eq:someConservationDiscrete2} to arrive at the discrete, control-volume-local formulation of~\cref{eq:someConservation},

\begin{equation}
  \sum_{\scv \in \cvSet_\cv} \frac{\someStorage_\scv^{t + \Delta t} - \someStorage_\scv^t}{\Delta t} + \sum_{\scvf \in \faceSet_\cv} \discFlux_{\cv, \scvf} - \sum_{\scv \in \cvSet_\cv} \someSource_\scv = 0,
  \label{eq:someConservationDiscrete3}
\end{equation}
where the approximate integrals $\someStorage_\scv$ and $\someSource_\scv$ depend on the specific finite volume method.
The time derivative is approximated by a backward or forward finite difference\footnote{\dumux currently only implements the backward and forward Euler time discretization schemes.}, determining the time level $t$ or $t + \Delta t$ at which the flux and source terms $\discFlux_{\cv, \scvf}$ and $\someSource$ are evaluated.
The expressions for the discrete fluxes $\discFlux_{\cv, \scvf}$ depend on the actual underlying finite volume scheme.
Fore more detailed information, we refer to \citep{HuberHelmig1999,helmig1997multiphase} for the \textsc{box} scheme,
and \citep{Droniou2014} for two-point and multi-point-flux-approximation finite volume schemes (\textsc{tpfa} and \textsc{mpfa}).

If a backward Euler time discretization is chosen, the nonlinear system corresponding to \cref{eq:someConservationDiscrete3} on each control volume $\cv$ is solved by Newton's method.
In each Newton step $n$, we assemble the discrete PDE system in residual form
\begin{equation}
A(u_n) \Delta u = A(u_n) (u_n - u_{n+1}) = r(u_n),
\end{equation}
where $u$ now is the vector of primary variables for each degree of freedom,
and $A = \frac{\partial r}{\partial u}$ is the Jacobian of the residual $r$, i.e.~the discrete equation evaluated at $u_n$.
The residual for a control volume $\cv\in\cvSet$ is given by the left-hand-side of~\cref{eq:someConservationDiscrete3}.
In the default configuration, the derivatives in the Jacobian matrix are approximated by numerical differentiation in \dumux.

\subsection{Element-wise assembly}
\begin{figure}[ht]
\centering
 \includegraphics[width=0.95\textwidth]{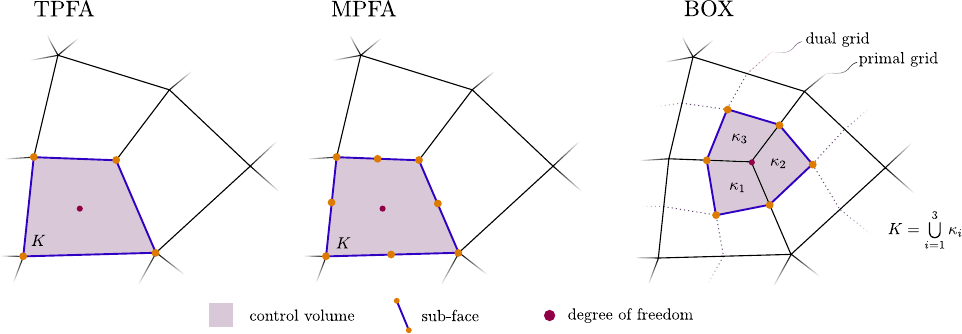}
\caption{Illustration of the configurations of control volumes and sub-faces on a computational grid for different discretization schemes.
The \textsc{tpfa} and \textsc{mpfa} scheme use the grid elements as control volumes ($\element\equiv\cv\equiv\scv$),
while in the \textsc{box} scheme the control volume is constructed around the grid vertices by connecting the barycenters of all adjacent geometrical entities.
The partition of the boundary $\partial \cv$ by the sub-faces $\scvf$ is illustrated by depicting the corners of each sub-face.}
\label{fig:controlVolumes}
\end{figure}

\dune grids and the connectivity information provided by the grid are rather element-centered (elements are grid entities of co-dimension $0$).
For this reason, it is natural in \dune to choose an element-wise assembly of the residual. The choice of $\cvSet$ and hence the
sub-control-volumes $\kappa$ and sub-control-volume-faces $\scvf$ is motivated by the goal of a convenient implementation
of an element-wise assembly for a given control volume scheme.
\Cref{fig:controlVolumes} depicts the configurations of control volumes and sub-faces as chosen in \dumux,
exemplarily for the \textsc{tpfa}, \textsc{mpfa} and the \textsc{box} scheme.
A corresponding illustration for the \textsc{mac} scheme can be found, for example, in \citep{schneider2019coupling}.
As illustrated in \cref{fig:controlVolumes}, the sub-control-volumes for the box scheme are
chosen such that the residual associated with a degree of freedom on a vertex
can be split into contributions from all neighboring elements.
For the \textsc{mpfa-o} scheme, the sub-faces are defined such that each sub-face can be associated
to a grid vertex. This is motivated by the fact that the expressions for the discrete
fluxes across the sub-faces depend on the degrees of freedom within so-called interaction
regions, constructed around the grid vertices \citep{Aavatsmark2002}.

\subsection{Representation in the software implementation}
Sub-control-volumes $\scv$ and sub-control-volume-faces $\scvf$, as introduced in~\cref{sec:fvmath},
are represented in the code by objects of the classes \cpp{SubControlVolume} and \cpp{SubControlVolumeFace}.
Depending on the caching model, which can be selected by specifying a template argument, these objects are either stored in the \cpp{GridGeometry} instance,
or constructed on-the-fly on a per-element basis during the assembly. Furthermore, \dumux has the notion of a \textit{local view}
on the grid geometry. The local view is an element-centered view that gives access to all $\scv$ and $\scvf$ on an element,
as well as all connectivity and geometry information to assemble storage, flux, and source terms. For example, for
a \textsc{tpfa} scheme, the neighboring control volumes and associated primary and secondary variables have to be accessed to
calculate the fluxes over the control-volume interfaces. \Cref{code:mass} shows how to assemble the total mass contained in a domain for a given grid $\mesh$.
The presented code is identical for all introduced finite volume schemes. Similar to the range-based-for syntax to iterate over
all elements of a grid view provided by \dune, the local view on the grid geometry in \dumux makes it possible to
iterate over all sub-control-volumes associated with an element in a concise and readable way (l.~17). Analogously, a range-generator
for iterating over all sub-control-volume-faces associated with an element is provided.

\begin{lstlisting}[style=cppstyle,basicstyle=\ttfamily\scriptsize,label={code:mass},
                   caption={The code to compute the total water mass in the domain $\Omega_h$
                            given a discrete water saturation field with values for each degree of freedom, stored in the vector \texttt{saturation}.
                            Most variables have been introduced in \cref{code:main}.}]
const double porosity = 0.4;
const double density = 1000.0;
double totalWaterMass = 0.0;

// iterate over all elements of the leaf grid view of the grid
for (const auto& element : elements(leafGridView))
{
    // construct a local view on the grid geometry
    auto fvGeometry = localView(*fvGridGeometry);

    // the view has to be bound to the current element
    // this is a no-op if the geometry is cached
    fvGeometry.bind(element);

    // iterate over all sub-control-volumes in the local view
    // and accumulate the mass
    for (const auto& scv : scvs(fvGeometry))
        totalWaterMass += saturation[scv.dofIndex()] * porosity
                          * density * scv.volume();
}

std::cout << "The total water mass is " << totalWaterMass << " kg/s" << std::endl;
\end{lstlisting}

Using the presented abstraction, the following discretization schemes have been implemented in \dumux:
\textsc{tpfa}, \textsc{mpfa-o}, \textsc{mpfa-l}, \textsc{mac} (staggered grid), finite volumes with non-linear two-point and multi-point flux approximation (\textsc{nl-tpfa}, \textsc{nl-mpfa}),
and mimetic finite differences (with additional degrees of freedom on the grid faces)~\citep{Schneider2018,Schneider2019},
from which \textsc{tpfa}, \textsc{box}, \textsc{mpfa-o}, and the \textsc{mac} scheme are available in the latest \dumux version.

\section{Multi-domain simulations}
\label{sec:multidomain}

\begin{figure}[ht]
\centering
 \includegraphics[width=1.0\textwidth]{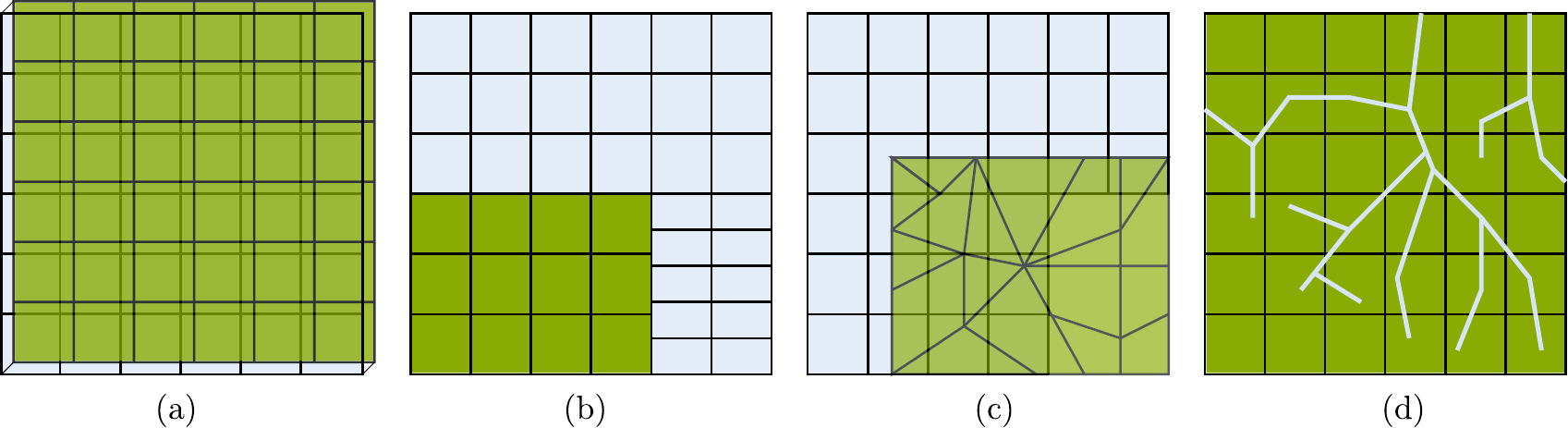}
\caption{Different types of model coupling in \dumux multi-domain simulations.
        (a)~multi-physics models (on the same grid),
        (b)~multiple non-overlapping domains with sharp conforming or non-conforming interface,
        (c)~multiple overlapping domains with different discretizations,
        (d)~conforming and non-conforming (embedded) mixed-dimensional domains (1D-2D, 1D-3D, 2D-3D).
        The different coupling modes can also be combined. Typical mixed-dimensional simulations also solve multi-physics problems,
        or use different discretization schemes in the subdomains. The number of subdomains is not limited to two.}
\label{fig:multidomainmodes}
\end{figure}

Model coupling is one of the key features in \dumux~\citep{Helmig2013,Koch2018a} and has been a driving force in the development of \dumux 3,
which features a new, consistent framework for implementing coupled models, that is, models composed of multiple coupled sub-models.
\Cref{fig:multidomainmodes} shows different types of model coupling that can be realized in this framework.
In this context, the term \textit{multi-domain simulation} covers a broad class of simulations featuring model coupling.
The idea is to solve a coupled PDE system, of which
a subset (a subdomain model) can be formulated on a different domain, potentially with different dimensionality,
and may be discretized using different meshes or different spatial discretization schemes.
The framework is designed to minimize software coupling, such that two existing \dumux models can be coupled without modifications
in the core components. To be minimally invasive, only the \cpp{Problem} class is required to be slightly modified to become a
\textit{sub-problem}. Sub-problems define their own initial and boundary conditions, and source term.
However, boundary and source terms may depend on quantities of other domains. For the data transfer, we introduce the concept
of a \textit{coupling manager} which is represented in the code by classes that implement the \cpp{CouplingManager} interface.
A coupling manager has to provide the information which degrees of freedom in one domain are coupled to which degrees of
freedom in another domain (the \textit{coupling stencil}). To this end, every coupling manager has to implement the function
shown in~\cref{code:coupstencil}.
\begin{lstlisting}[style=cppstyle,basicstyle=\ttfamily\scriptsize,label={code:coupstencil},
                   caption={The member function \texttt{couplingStencil} has to be implemented by all deriving coupling manager implementations.
                            The template arguments \texttt{i} and \texttt{j} are the indices of one pair of coupled subdomains. They are deduced
                            from the two index objects \texttt{domainI} and \texttt{domainJ}, passed as arguments to the function.
                            In case two subdomain are not coupled, the function is required to return a reference to an empty stencil vector.
                            For domain \texttt{i}, an instance of the element is passed, for which the residual in the
                            element-wise assembly is to be computed. The function returns a vector of all indices of degrees of freedom
                            coupled to one of the degrees of freedom associated with element \texttt{elementI}.}]
// required member function of every CouplingManager class
// return the coupling stencil for element elementI (domain i)
// with respect to degrees of freedom in domain j
template<std::size_t i, std::size_t j>
const std::vector<std::size_t>&
couplingStencil(Dune::index_constant<i> domainI,
                const Element<i>& elementI,
                Dune::index_constant<j> domainJ) const;
\end{lstlisting}
Moreover, the coupling manager has to transfer data between the sub-problems.
This general concept can be used to implement a wide class of coupling schemes. Finding the coupling stencils often involves
intersecting two grids. For this purpose, \dumux provides efficient intersection algorithms
based on axis-aligned bounding box volume hierarchy data structures~\citep{ericson2004real,Massing2013} computed for \dune grids.
\Cref{code:intersection} shows how to intersect two given grid geometries and use the connectivity information to construct a
coupling stencil, that can be used, for example,
in a non-conforming embedded fracture model\footnote{The \dumux code for such a simulation (2D-3D embedded fracture model) can be found in the \dumux repository under
\href{https://git.iws.uni-stuttgart.de/dumux-repositories/dumux/tree/master/test/multidomain/embedded/2d3d/1p_1p}{\url{test/multidomain/embedded/2d3d/1p_1p}}}.
\begin{lstlisting}[style=cppstyle,basicstyle=\ttfamily\scriptsize,label={code:intersection},
                   caption={Sample code which intersects two grid geometries (see~\cref{sec:fvmath}) and uses the resulting connectivity information to construct coupling stencils for a cell-centered finite volume scheme.
                            A bulk grid (\texttt{bulk}) is intersected with an overlapping grid with lower dimension (\texttt{lowDim}) which may
                            discretize a three-dimensional rock domain and a network of two-dimensional fractures. (The code snippet is identical for 3D-2D and 2D-1D.) An intersection is $\Gamma = E_b \cap E_l$,
                            where $E_b$ (\texttt{targetEntity}) is an element of the bulk grid and $E_l$ (\texttt{domainEntity}) an element of the lower-dimensional embedded grid. If there
                            exist multiple such intersections with identical geometry (e.g. a fracture element coincides with the intersection of two bulk elements),
                            they are represented in the code by a single intersection object with access to all elements ("entities") associated with the intersection ("neighbors").
                            The interface of the \texttt{glue} object is similar to that implemented in the \dune module \dunemodule{dune-grid-glue}~\citep{Bastian2010}
                            which implements an advancing front algorithm instead of an algorithm based on spatial data structures used here.}]

// which degrees of freedom of the other domain does the element residual depend on?
using Stencils = std::size_t<std::set<std::size_t>>;
Stencils lowDimCouplingStencils, bulkCouplingStencils;
// resize to the number of grid elements
lowDimCouplingStencils.resize(lowDimGridGeometry.gridView().size(/*codim=*/0));
bulkCouplingStencils.resize(bulkGridGeometry.gridView().size(/*codim=*/0));

// intersect two grid geometries
// we choose domain (first argument): lowDim, target (second argument): bulk
// see <dumux/multidomain/glue.hh>
const auto glue = makeGlue(lowDimGridGeometry, bulkGridGeometry);

// interate over all intersections
for (const auto& is : intersections(glue))
{
    // the element index of the lowDim element of this intersection
    const auto domainIdx = lowDimGridGeometry.elementMapper().index(is.domainEntity(0));
    // there might be multiple bulk elements associated with this intersection
    for (unsigned int i = 0; i < is.numTargetNeighbors(); ++i)
    {
       // the element index of the bulk element of this intersection
        const auto targetIdx = bulkGridGeometry.elementMapper().index(is.targetEntity(i));
        // insert target-domain index pair into the respective stencil
        lowDimCouplingStencils[domainIdx].insert(targetIdx);
        bulkCouplingStencils[targetIdx].insert(domainIdx);
    }
}
\end{lstlisting}

Furthermore, \dumux provides an assembler class for multi-domain models, which assembles the discrete PDE system in residual form
\begin{equation}
\begin{bmatrix}
A_1 &  C_{12} & \dots & C_{1n} \\
C_{21} &  A_2 & & \\
\vdots &  & \ddots & \\
C_{n1} &  & &  A_n
\end{bmatrix}
\begin{bmatrix}
 \Delta u_1 \\
 \Delta u_2 \\
 \vdots\\
 \Delta u_n
\end{bmatrix}
=
\begin{bmatrix}
r_1 \\
r_2 \\
\vdots \\
r_n \\
\end{bmatrix},
\end{equation}
where $A_i$ is the Jacobian of the discrete PDE system for subdomain $i$,
and $C_{ij}$ is the coupling Jacobian with derivatives of residuals of domain $i$ with respect to degrees of freedom
of domain $j$, $C_{ij} = \frac{\partial r_i}{\partial u_j}$. The assembler class and the matrix class are generic,
so that the sub-vectors $u_i$ and sub-matrices $A_i$ can themselves have a block structure,
and support an arbitrary number of subdomains.
The block structure can be exploited, for example, for constructing preconditioners for a monolithic solver, or to
algebraically implement schemes where the subdomain systems are solved successively in an iterative algorithm.

The presented multi-domain concept has been successfully used to implement models with coupled flow and transport processes
in vascularized brain tissue~\citep{Vidotto2018,Koch2018b}, root--soil interaction models in the vadose zone~\citep{Koch2018a},
flow and transport models for fractured rock systems~\citep{glaser2017discrete,glaser2019hybrid},
coupled porous medium flow and atmosphere flow (Darcy-Navier-Stokes)
at the soil surface~\citep{schneider2019coupling},
and a model that couples a pore-network model with a Navier-Stokes model~\citep{weishaupt2019a}.

\section{New features in \dumux 3}
\label{sec:features}

Numerous new features are added in \dumux 3 in comparison with the 2.X series. We briefly mention the most important changes.
A complete redesign of many high-level class abstractions such as assembler, linear and non-linear solvers,
grid readers and grid geometry, and file I/O leads to more readable and flexible \cpp{main} functions, see also~\cref{sec:design}.
The numerous models are improved in terms of code reuse and modularity, so that code duplication is minimized and readability improved.
In addition to the fluid system concept, a solid system concept is introduced, which facilitated the implementation of
new models including mineralization or precipitation of substances that potentially modify the porous matrix structure~\citep{Cunningham2019,Hommel2018}.
Porous material properties such as intrinsic permeability and porosity can be implemented to depend (linearly or non-linearly) on the primary variables.
We generalized thermal and chemical non-equilibrium models to be combinable with any porous medium model. Multi-component
diffusion can now be modeled by Maxwell-Stefan diffusion.
A versatile implementation of cell-centered \textsc{mpfa-o} scheme is now usable with all models~\citep{glaser2017discrete}.
The Navier-Stokes models are redesigned to use a \textsc{mac} scheme on a staggered grid~\citep{fetzer2017,schneider2019coupling,weishaupt2019a},
including a Reynolds-averaged Navier-Stokes model~\citep{Yang2019,Fetzer2016}
with a variety of turbulence models (e.g.~$k$-$\epsilon$, $k$-$\omega$), as well as a second-order upwind scheme.
We can now solve problems based on the two-dimensional shallow water equations.
Finally, the multi-domain module adds the functionality of versatile model coupling as described in~\cref{sec:multidomain}.

\section{Numerical examples}
\label{sec:gallery}
In the following section, three numerical examples demonstrate the flexibility of the multi-domain framework in \dumux 3.
In~\cref{sec:ex_ffpm}, free flow over a porous medium is modeled by coupling a Navier-Stokes model to a pore-network model. The domains are coupled at a common interface.
\Cref{sec:ex_frac} shows a simulation of two-phase flow in a fractured rock matrix. The fracture flow is computed
on lower-dimensional domains conforming with the control-volume faces of the three-dimensional rock matrix domain discretization,
which allows to model highly conductive fractures as well as impermeable fractures.
The example shows the differences between a \textsc{tpfa} and an \textsc{mpfa-o} finite volume scheme
for a rock matrix with anisotropic permeability.
Finally, \label{sec:ex_root} an example of root water uptake and tracer transport is given. The roots are
represented by a network of tubes embedded into the soil matrix. The mass exchange between the two non-conforming domains
is realized with adequate source terms. The source code for all examples as well as instructions for the reproduction of the results can be found in the \dumux-pub module
to this publication at \href{https://git.iws.uni-stuttgart.de/dumux-pub/dumux2019}{\url{git.iws.uni-stuttgart.de/dumux-pub/dumux2019}}.

\subsection{Coupling a free flow model with a pore-network model}
\label{sec:ex_ffpm}
This example is adapted from~\citep{weishaupt2019a}, where a coupled model of free channel flow adjacent to a porous medium is presented
with a detailed model description. We model transient multi-component flow over a random porous structure in a two-dimensional model domain.
The channel flow is governed by the Navier-Stokes equations, neglecting gravity and dilatation~\citep{truckenbrodt1996a},
\begin{equation}
\label{eq:NavierStokes}
 \frac{\partial (\rho \mathbf{v})}{\partial t} + \nabla \cdot \left(\rho \mathbf{v} \mathbf{v}^T \right) = \div{\left[\mu \left(\grad{ \mathbf{v}} + \gradt{ \mathbf{v}}\right)\right]} -\grad{p},
\end{equation}
with the fluid density $\rho$, the fluid velocity $\mathbf{v}$, the fluid pressure $p$, and the dynamic viscosity $\mu$.
A molar balance equation for each component $\kappa$,
\begin{equation}
\label{eq:ComponentBalance}
  \frac{\partial \left(\rho_\text{m} x^\kappa\right)}{\partial t}
    + \nabla \cdot \left( \rho_\text{m} x^\kappa {\mathbf{v}}
   - D  \rho_\text{m} \grad{x^\kappa} \right)
    = 0,
\end{equation}
models advective and diffusive transport of the fluid components, where
$x^\kappa$ is the component mole fraction, $\rho_\text{m}$ the fluid molar density, and $D$ the binary diffusion coefficient. Diffusive fluxes
are described by Fick's law.

For modeling flow and transport in the porous medium, a pore-network model is used \cite{fatt1956a, blunt2017a}.
The complex pore-scale geometry of the porous medium is
transferred to a simplified geometry, the pore bodies and pore throats. For each component $\kappa$,
\cref{eq:ComponentBalance} is formulated discretely on each pore body where the primary variables are located.
The advective and diffusive fluxes are evaluated on the pore throats.
We refer to \citep{weishaupt2019a} for further details.

Coupling conditions are formulated at the interface between the two models to ensure thermodynamic consistency. At the locations where no throat
intersects with the interface, a no-flow/no-slip condition for the Navier-Stokes model is enforced.
At the actual intersections, we prescribe the continuity of normal forces \cite{layton2002a} resulting in a Neumann boundary condition for the free flow domain,
\begin{equation}
\mathbf{n} \cdot \left[\left(\left(\rho \mathbf{v} \mathbf{v}^T -\mu \left(\grad{ \mathbf{v}} + \gradt{ \mathbf{v}}\right) + p\mathbf{I}\right) \mathbf{n}\right)\right]^{\textsc{ff}}  = [p]^{\textsc{pnm}},
\end{equation}
where the superscripts \textsc{ff} and \textsc{pnm} mark quantities of the free-flow and pore-network model, respectively.
We use the tangential part of the average velocity within the throat at the boundary $[ \mathbf{v} ]^{\textsc{pnm}}$ as a slip condition
for the free flow,
\begin{equation}
 [\mathbf{v} \cdot \mathbf{t}]^{\textsc{ff}} = \begin{cases}
 [ \mathbf{v} ]^{\textsc{pnm}} \cdot [\mathbf{t}]^{\textsc{ff}} \quad \text{on pore throat},\\
0 \hfill\text{else},
\end{cases}
\end{equation}
where $[\mathbf{t}]^{\textsc{ff}}$ is a unit tangential vector to the coupling interface.

The velocity within the throat is given by
\begin{equation}
 [\mathbf{v} ]^{\textsc{pnm}} = \frac{Q_t}{A_t} \mathbf{n_t},
\end{equation}
where $Q_{t}$ is the volume flow within the throat, $A_{t}$ is its respective cross-sectional area and $\mathbf{n_{t}}$ is a unit vector parallel to the pore throat, pointing towards the coupling interface.
Finally, we require the conservation of mass
\begin{equation} \label{eq:CouplingMolarFlux}
 [(\rho_\text{m} x^\kappa \mathbf{v} -
 D \rho_\text{m} \nabla x^\kappa) \cdot \mathbf{n}]^{\textsc{ff}} = -[(\rho_\text{m} x^\kappa \mathbf{v} -
 D \rho_\text{m} \nabla x^\kappa) \cdot \mathbf{n}]^{\textsc{pnm}}
\end{equation}
and enforce the continuity of mole fractions at the interface,
\begin{equation}
 [x^\kappa]^{\textsc{ff}} = [x^\kappa]^{\textsc{pnm}}.
\end{equation}

The Navier-Stokes equations are discretized with a \textsc{mac} scheme on a staggered grid.
The pore-network model is also implemented in \dumux and will become part of stable code basis in an upcoming release.
The pore-network is described as a one-dimensional network embedded in a two-dimensional domain, using the \dune grid implementation \dunemodule{dune-foamgrid}~\citep{foamgrid}.
We used the \dumux multi-domain framework in order to achieve
a fully monolithic coupling between the two sub-models.
Since only elements on the domain boundary are coupled in this example, we employed a simplified
intersection algorithm for creating the coupling stencils in comparison with \cref{code:intersection}. Instead of intersecting the entire grid,
only the end points (boundary pores) of the pore-network grid geometry are intersected with the channel domain grid geometry to compute the coupling stencils.
Furthermore, for the staggered grid discretization, we compute separate coupling stencils for the degrees of freedom located at cell centers and those located
on cell faces. Newton's method is used to solve the nonlinear system of equations in combination with SuiteSparse's UMFPack~\citep{umfpack2004} as a direct
linear solver. Implementation details can be found in the folder \texttt{dumux/multidomain/boundary}
located in the \dumux repository and in the \dumux-pub module accompanying this paper (see above).

\begin{figure}[ht]
\centering
 \includegraphics[scale=0.04]{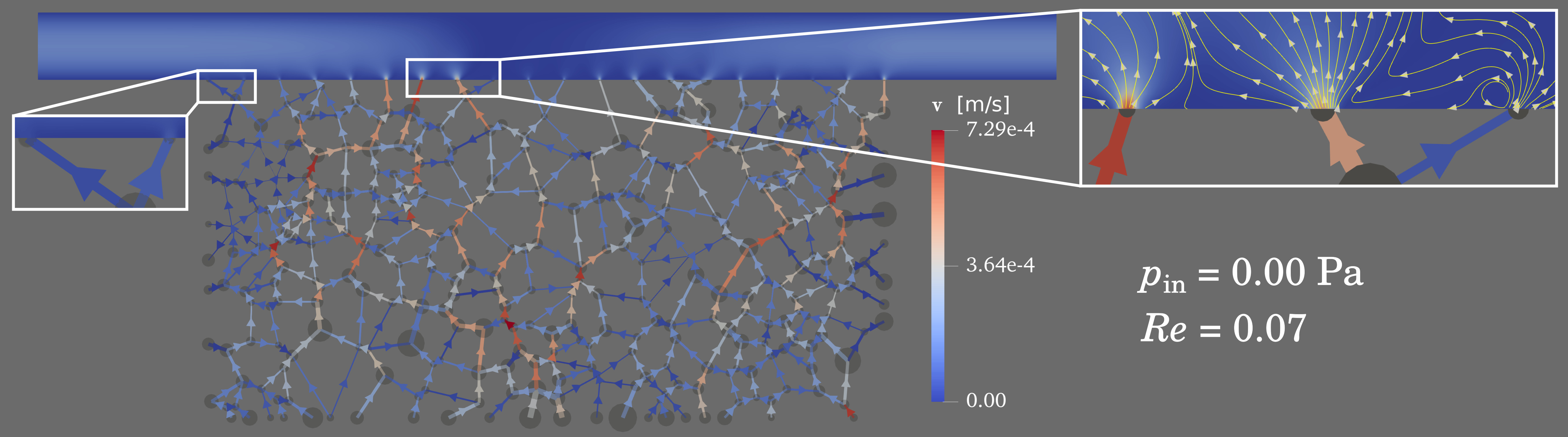} \\
 \vspace{0.1cm}
 \includegraphics[scale=0.04]{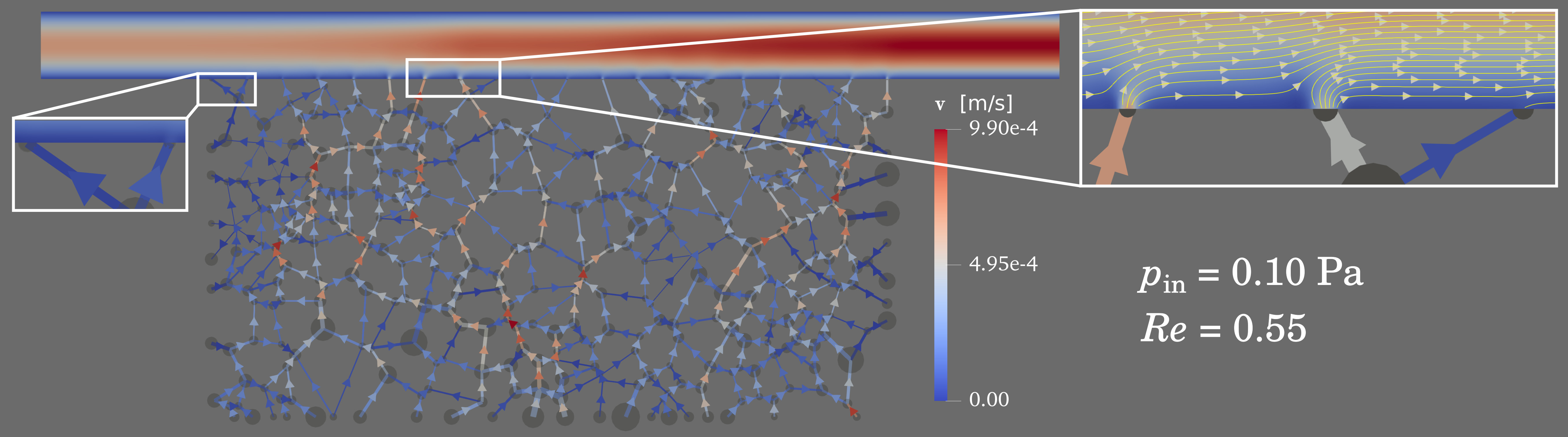} \\
 \vspace{0.1cm}
 \includegraphics[scale=0.04]{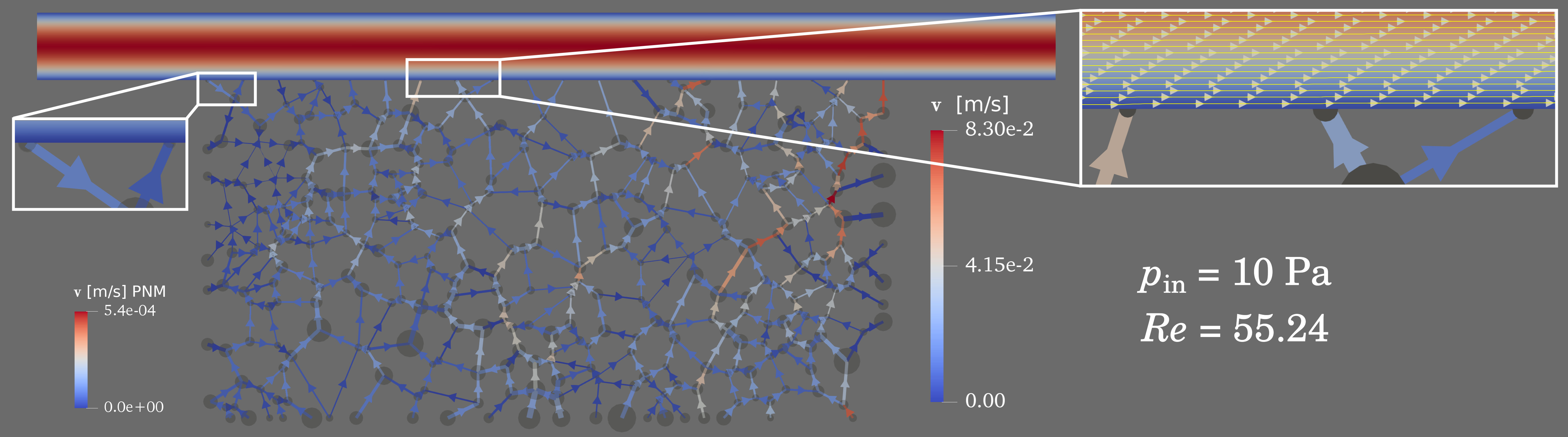}
\caption{Velocity fields for the three scenarios. $Re$ is based on the averaged velocity within the channel. Note the different color scale for the network in the third scenario (bottom). $p_\mathrm{out} = 0$.
  Figure adapted from~\citep{weishaupt2019a} (license: CC BY 4.0).}
\label{fig:voronoi_v}
\end{figure}

\begin{figure}[ht]
\centering
\begin{subfigure}{.49\textwidth}
  \captionsetup{skip=1pt, justification=centering, font=scriptsize}
  \centering
  \includegraphics[scale=0.05]{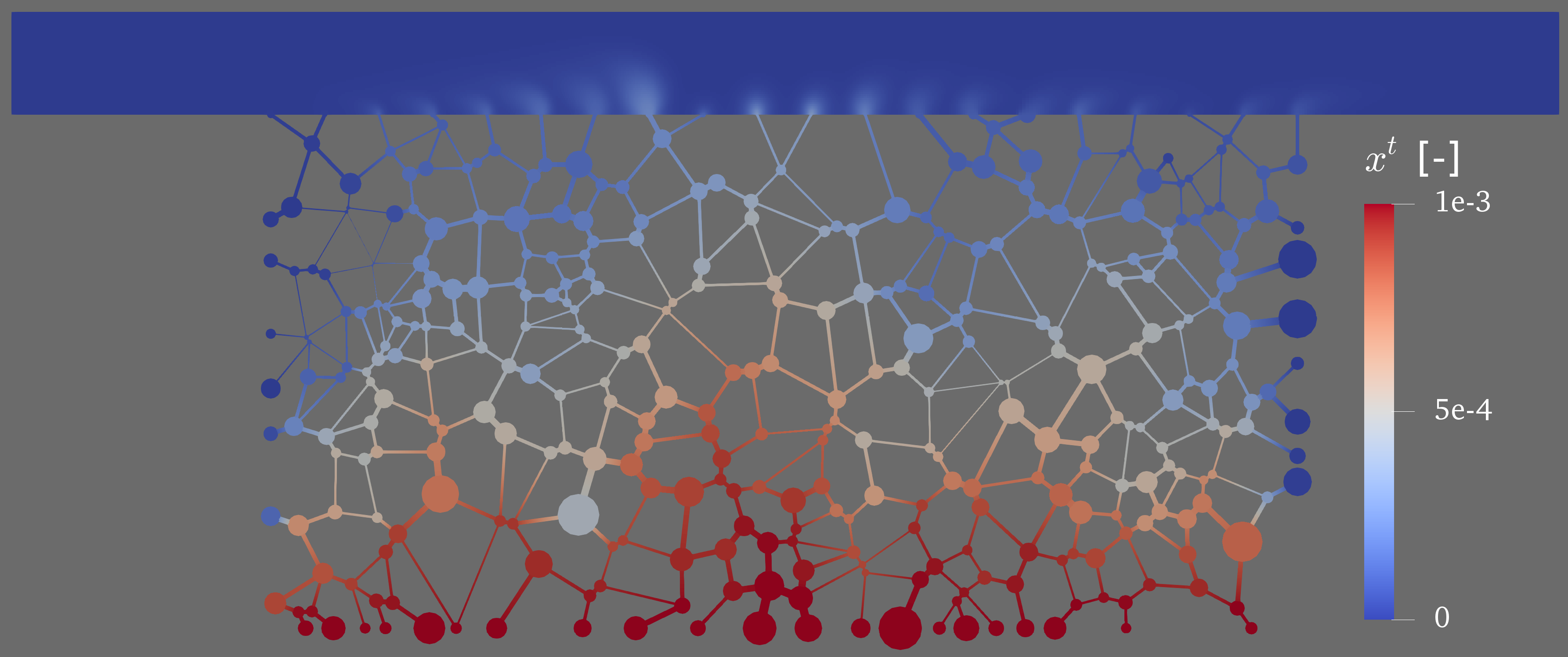}
  \caption*{$p_{in} = \SI{0}{\pascal} $, $Re = \SI{0.07}{}$, $t_1 = \SI{14}{\second}$}
  \vspace{0.5ex}
\end{subfigure}
\begin{subfigure}{.49\textwidth}
  \captionsetup{skip=1pt, justification=centering, font=scriptsize}
  \centering
  \includegraphics[scale=0.05]{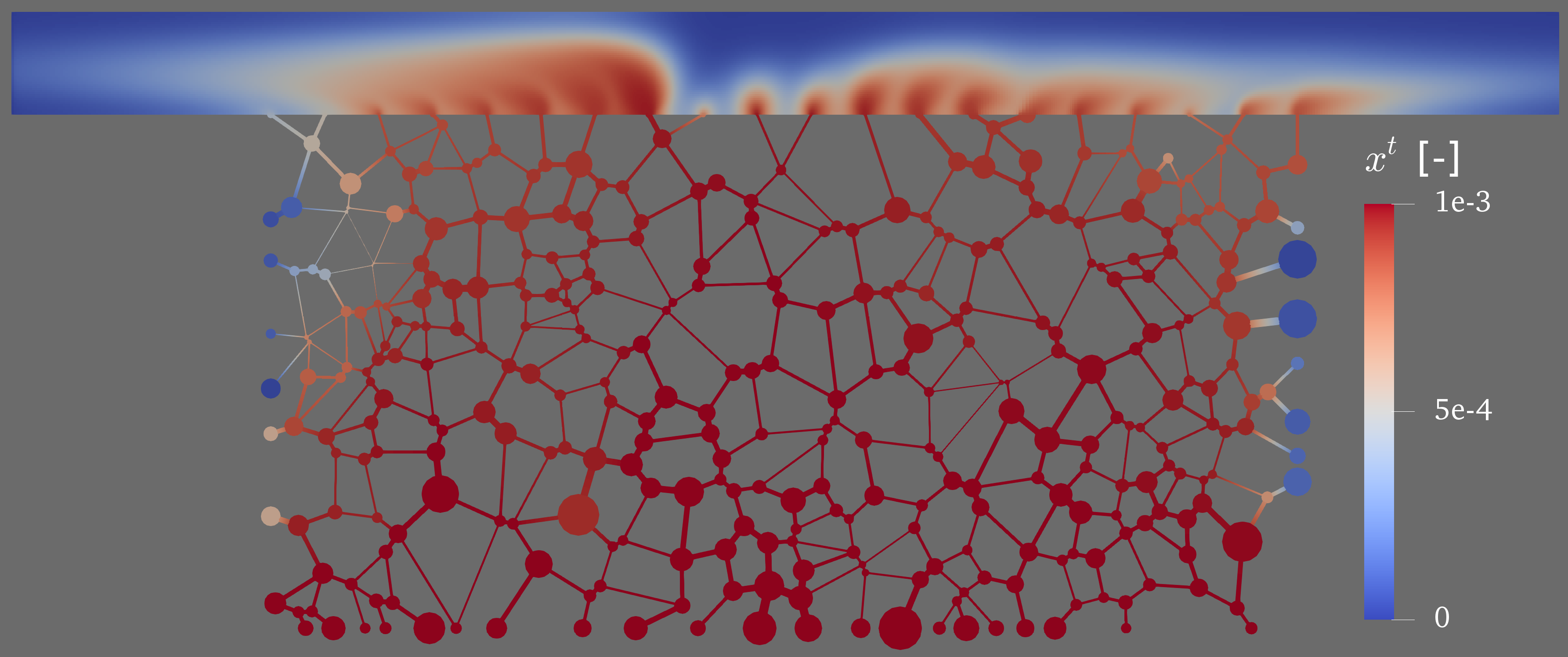}
  \caption*{$t_2 = \SI{50}{\second}$}
  \vspace{0.5ex}
\end{subfigure}
\begin{subfigure}{.49\textwidth}
  \captionsetup{skip=1pt, justification=centering, font=scriptsize}
  \centering
  \includegraphics[scale=0.05]{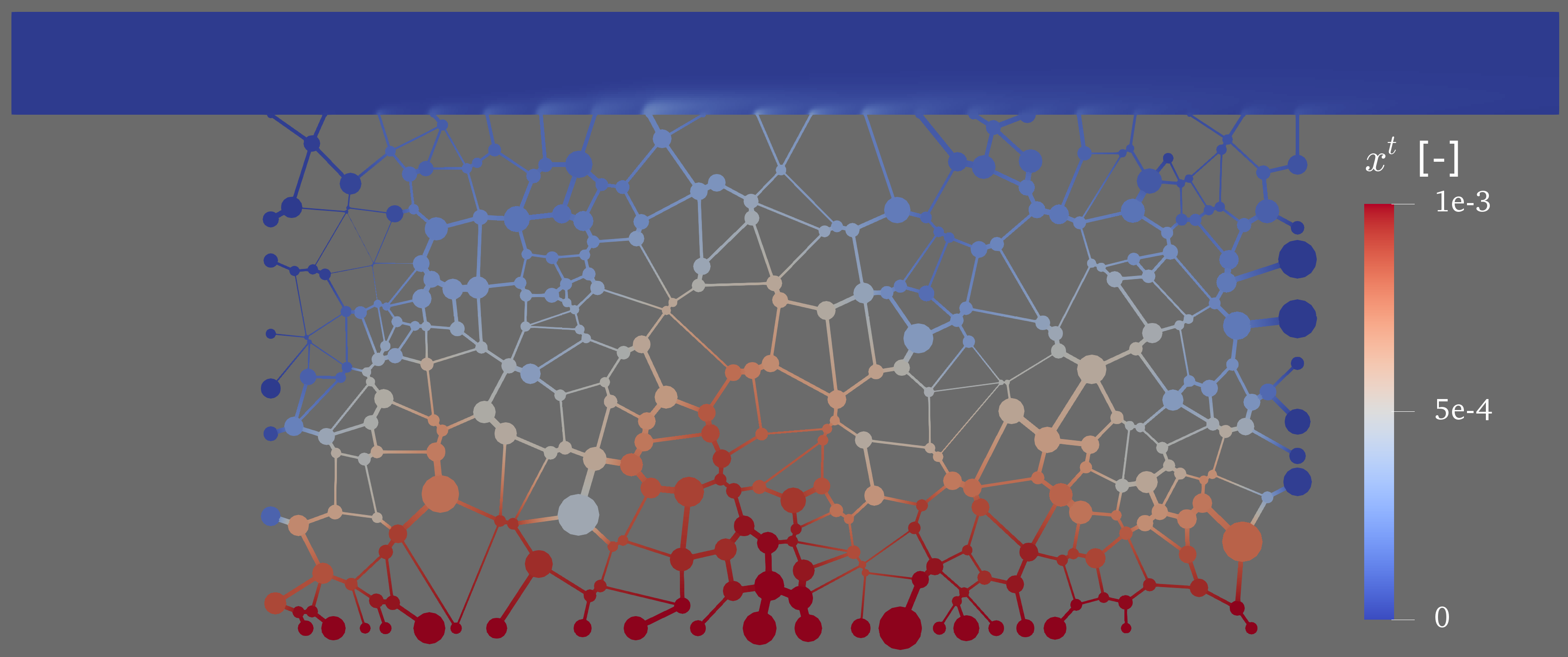}
  \caption*{$p_{in} = \SI{0.1}{\pascal} $, $Re = \SI{0.55}{}$, $t_1 = \SI{14}{\second}$}
  \vspace{0.5ex}
\end{subfigure}
\begin{subfigure}{.49\textwidth}
  \captionsetup{skip=1pt, justification=centering, font=scriptsize}
  \centering
  \includegraphics[scale=0.05]{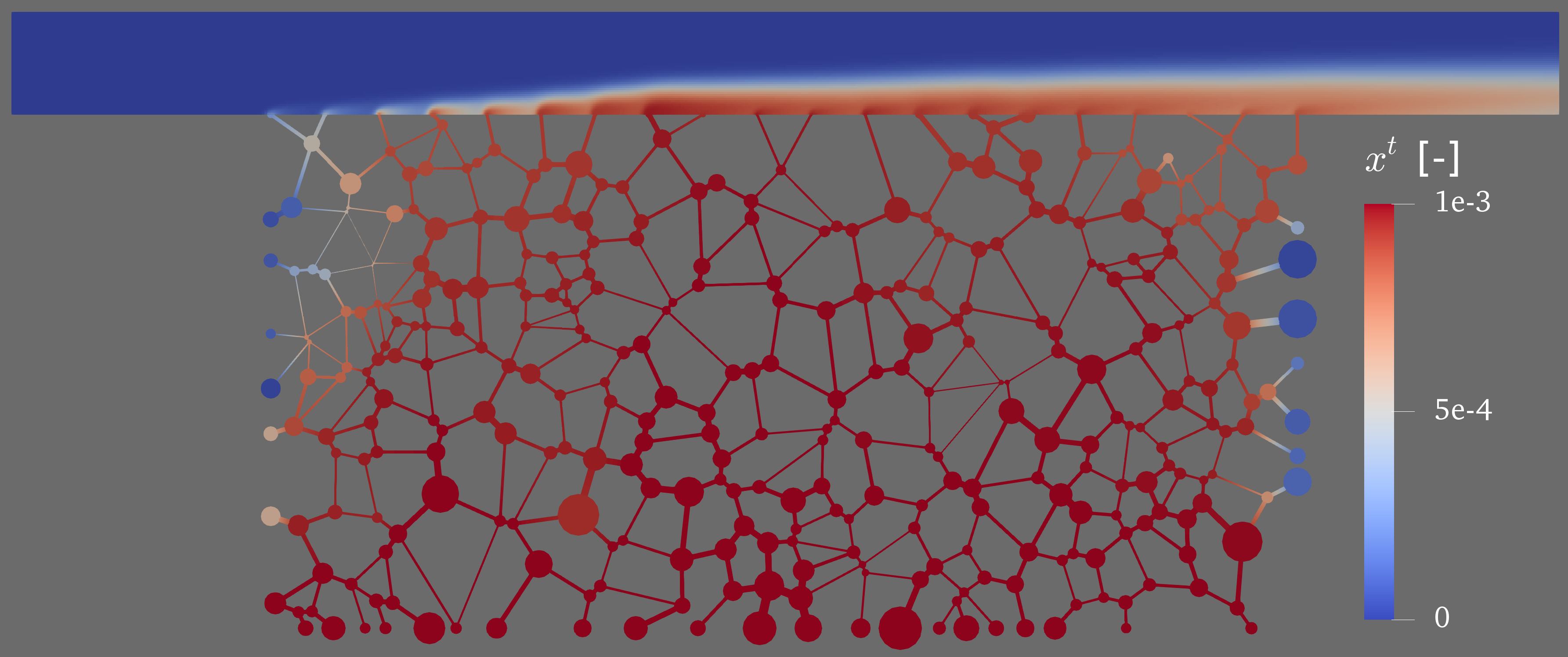}
  \caption*{$t_2 = \SI{50}{\second}$}
  \vspace{0.5ex}
\end{subfigure}
\begin{subfigure}{.49\textwidth}
  \captionsetup{skip=1pt, justification=centering, font=scriptsize}
  \centering
  \includegraphics[scale=0.05]{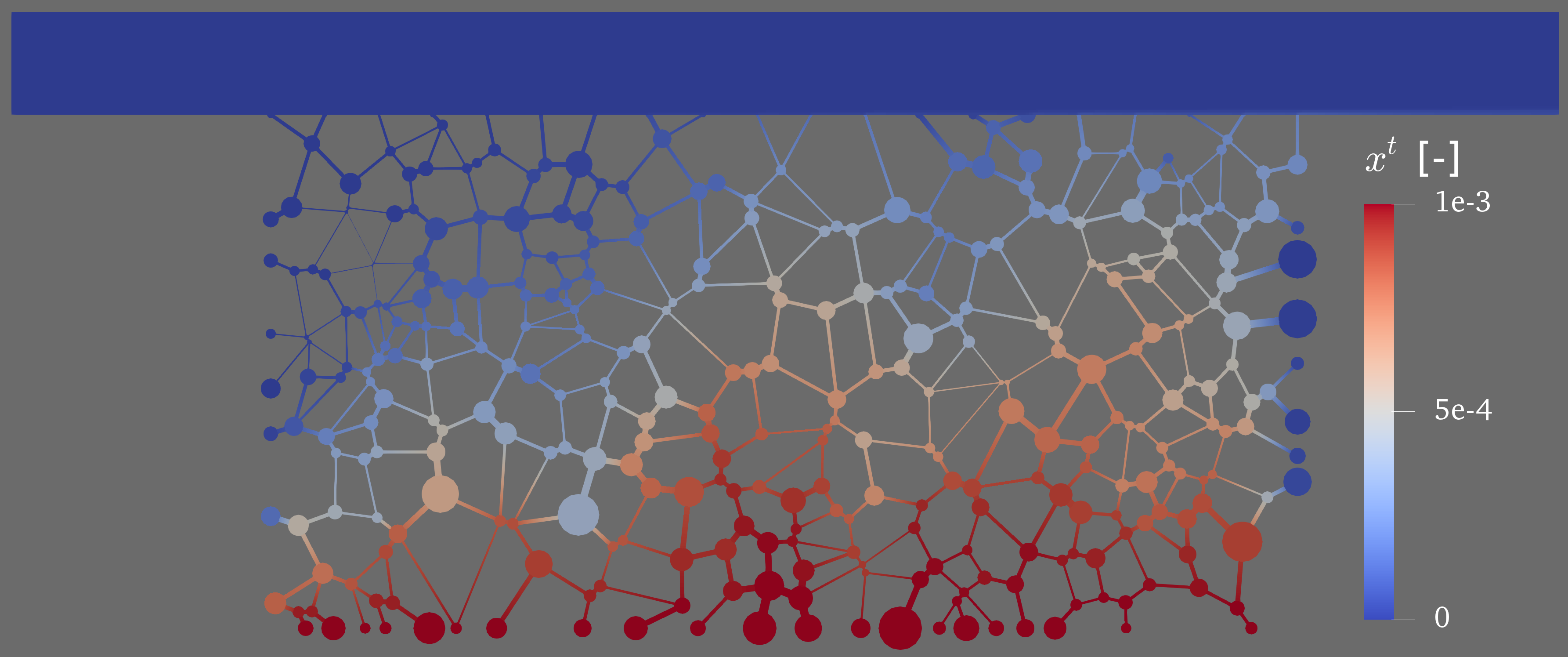}
  \caption*{$p_{in} = \SI{10}{\pascal} $, $Re = \SI{55.24}{}$, $t_1 = \SI{28}{\second}$}
  \vspace{0.5ex}
\end{subfigure}
\begin{subfigure}{.49\textwidth}
  \captionsetup{skip=1pt, justification=centering, font=scriptsize}
  \centering
  \includegraphics[scale=0.05]{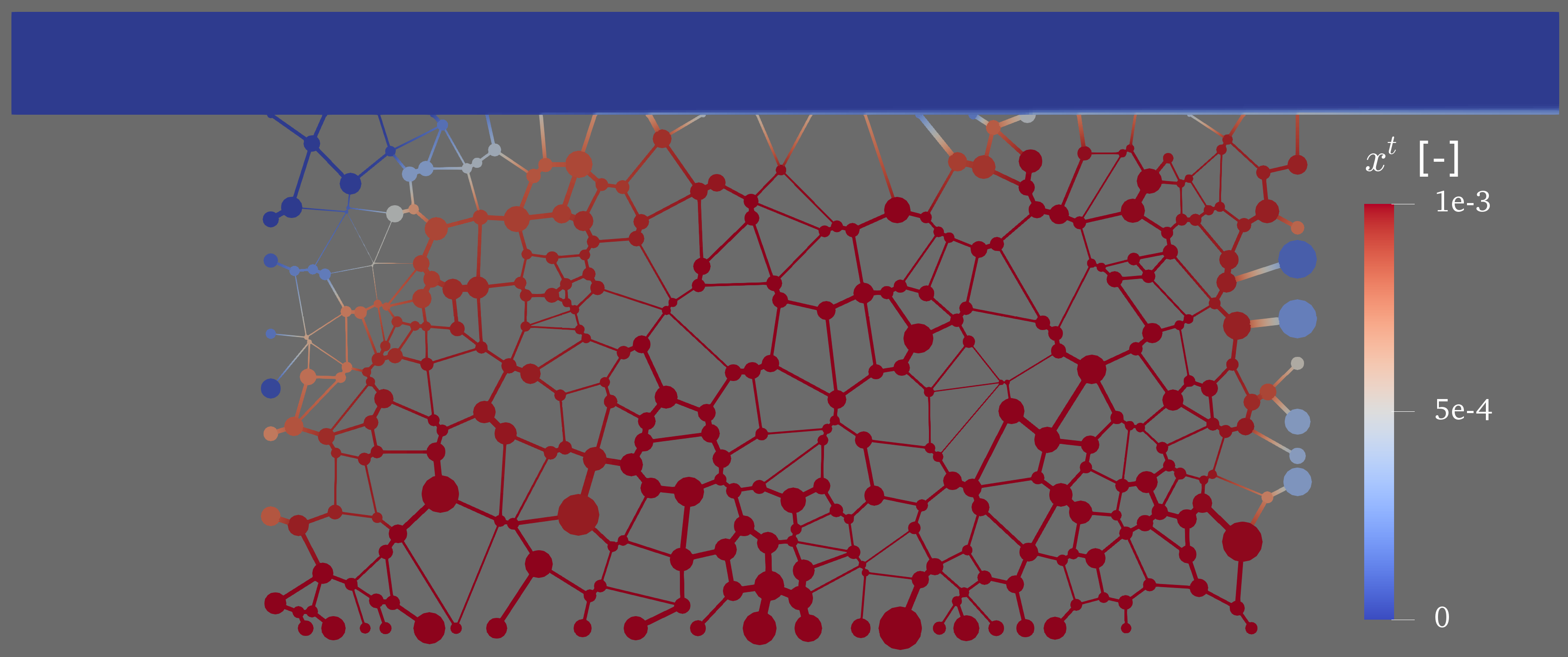}
  \caption*{$t_2 = \SI{126}{\second}$}
  \vspace{0.5ex}
\end{subfigure}
\caption{Distribution of the mole fraction $x^t$ for the three different scenarios at different times. Figure adapted from~\citep{weishaupt2019a} (license: CC BY 4.0).}
\label{fig:voronoi_time}
\end{figure}

The given example is discussed in detail in~\citep{weishaupt2019a}. Fluid density and viscosity are assumed constant, with $\rho = \SI{1e3}{kg/m^3}$ and $\mu = \SI{1e-3}{\pascal \second}$. A tracer injected at the bottom of the pore network is transported
upwards until it reaches the free-flow channel through which it leaves the system at its left or right sides where
fixed pressures are set (see~\cref{fig:voronoi_v}).
All other sides of the channel are closed and no-flow/no-slip conditions hold. In the pore network, Dirichlet conditions for $p$ and $x^\kappa$ are set at the bottom while Neumann
no-flow boundaries are assigned to the lateral sides. Varying the pressure gradients in the free-flow channel yields three different scenarios.
\Cref{fig:voronoi_v} shows the resulting velocity fields where distinct preferential flow paths in the network
and the influence of the inclined throats at the interface become visible. \Cref{fig:voronoi_time} shows the temporal development of the
concentration fields. At $t_1$, an average mole fraction of $\SI{5e-4}{}$ is reached while $t_2$ corresponds to a value of $\SI{9e-4}{}$.
While the first two cases reach these points after 14 and 50 seconds, the higher pressure gradient in the channel for the case with $Re = 55.24$
repels the tracer, keeping it longer in the network. As there is no imposed flow for the first case, the tracer spreads equally to the left and right side of the channel while in the other
two cases, the pressure gradient drives the tracer towards the right outlet. The
formation of a boundary layer at the interface can be observed which becomes thinner for the higher $Re$ case.

\subsection{Two-phase flow through fractured porous media}
\label{sec:ex_frac}

The example shown in this section is inspired by an exercise of the \dumux-course\footnote{The exercise can be found at \href{https://git.iws.uni-stuttgart.de/dumux-repositories/dumux-course/tree/master/exercises/exercise-fractures}{\url{git.iws.uni-stuttgart.de/dumux-repositories/dumux-course/tree/master/exercises/exercise-fractures}}}
and considers the buoyancy-driven upwards migration of gaseous nitrogen in an initially fully water-saturated fractured porous medium.
In this model, the fractures are assumed to represent very thin heterogeneities with substantially differing material parameters in comparison to the surrounding porous medium.
The thin nature of these inclusions favors a dimension-reduced description of the fractures as entities of co-dimension $1$, on which cross-section averaged PDEs are solved
and appropriate coupling conditions describe the interaction with the surrounding porous medium.
Such approaches have been widely reported in the literature for both single-phase flow (see e.g.~\citep{martin2005modeling,Ahmed2015MpfaDfm,flemisch2018Benchmarks})
and two-phase flow (see e.g.~\citep{Reichenberger2006mixed,Fumagalli2013DfmTwoPhase,Tene2017pEDFM}) in fractured porous media.
In the approach presented in this example, the two subdomains are not discretized independently, but it requires the facets of the higher-dimensional grid (bulk grid) to be conforming with the elements of the lower-dimensional grid for the fractures.
Therefore, we currently rely on grid file formats that allow the extraction of both grids from a single file together with the connectivity information between them.
The implementation in this example uses mesh files generated by Gmsh~\citep{Gmsh2009}.
This facilitates the determination of the coupling stencils, making it obsolete to intersect the grids, in contrast to the procedure presented in~\cref{code:intersection} for non-conforming methods.
All classes and functions related to this approach can be found in the \texttt{dumux/multidomain/facet} folder of the \dumux repository.
For further details on the numerical scheme, we refer to~\citep{glaser2017discrete,glaser2019hybrid}.

\begin{figure}[ht]
\centering
 \includegraphics[width=0.75\textwidth]{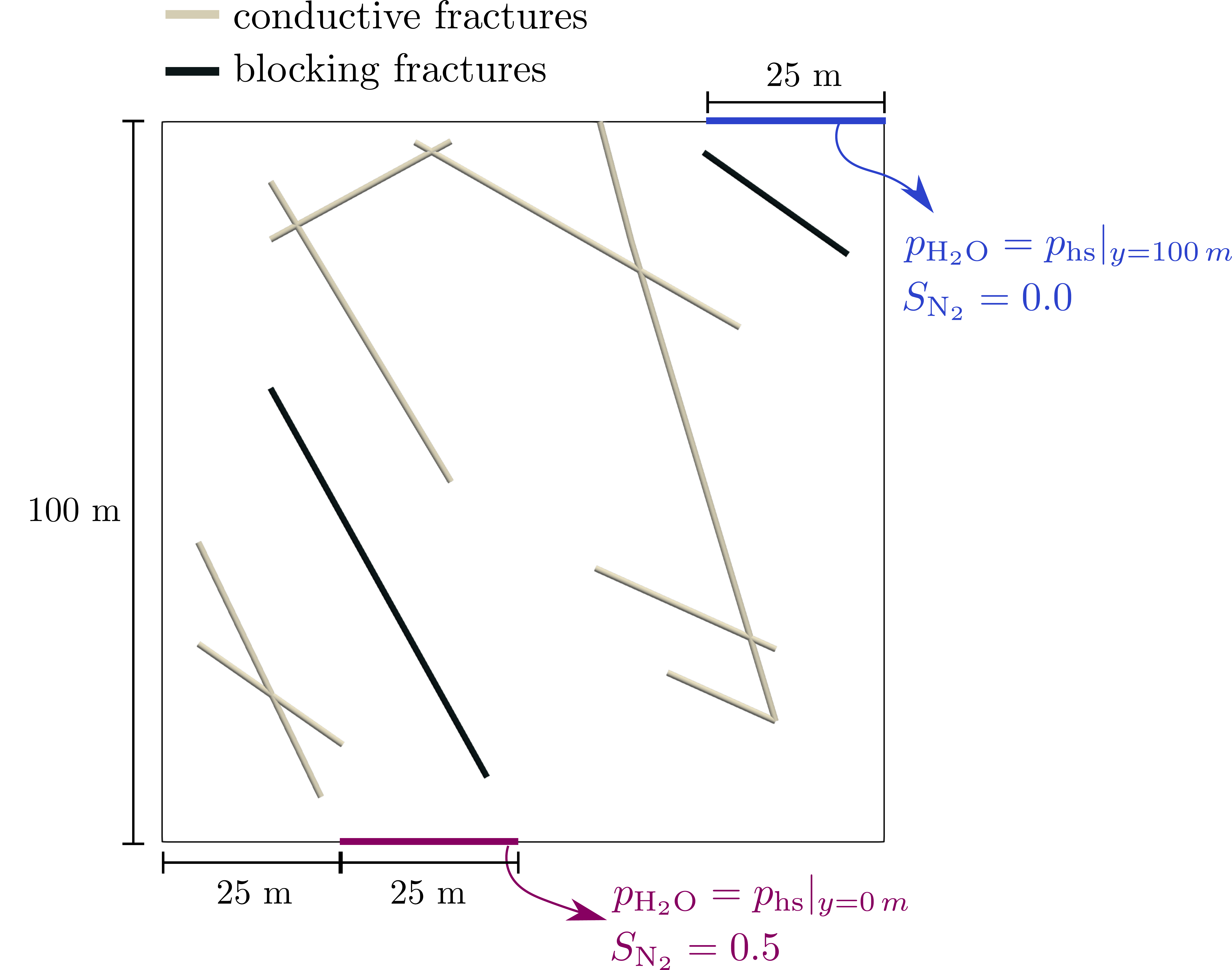}
\caption{Domain and Dirichlet boundary conditions for the two-phase flow example through fractured porous media. The subscript $\mathrm{hs}$ refers to hydrostatic pressure conditions.}
\label{fig:fractureExampleDomain}
\end{figure}

The fracture network geometry used in this example is taken from~\citep{flemisch2018Benchmarks} and is shown together with the boundary conditions in~\cref{fig:fractureExampleDomain}.
The computational domain comprises a two-dimensional domain for the matrix, $\Omega_\text{m}$ and a one-dimensional domain for the fracture network, $\Omega_\text{f}$.
We denote by $\Gamma_\text{m}$ the outer boundary of $\Omega_\text{m}$ and further decompose this into $\Gamma_\text{m}^\text{D}$ and $\Gamma_\text{m}^\text{N}$, referring to the subsets on which Dirichlet and Neumann conditions are specified, respectively.
The boundary of $\Omega_\text{f}$ is denoted with $\Gamma_\text{f}$. Furthermore, we introduce the partitions $\Omega_\text{f}^\text{C}$ and $\Omega_\text{f}^\text{B}$ such that $\Omega_\text{f} = \Omega_\text{f}^\text{C} \cup \Omega_\text{f}^\text{B}$, which refer to the conductive (superscript C) and blocking fractures (superscript B).
Finally, we define the interfaces between the matrix domain and the conductive and blocking fractures as $\gamma_\text{C} = \Omega_\text{m} \cap \Omega_\text{f}^\text{C}$ and $\gamma_\text{B} = \Omega_\text{m} \cap \Omega_\text{f}^\text{B}$.
Making use of this notation, we state the governing system of equations as follows:

\begin{subequations}
  \label{eq:ex_frac_problem}
  \begin{align}
    \darcyVelocity_{\fluidPhaseIdx, i}
      + \frac{\relPerm{\fluidPhaseIdx, i}}{\dynVisc_\fluidPhaseIdx} \permeability_i
        \left( \grad \pressure_{\fluidPhaseIdx, i} - \density_{\fluidPhaseIdx} \gravity \right)
    &= 0, \label{eq:ex_frac_darcy}\\
    \frac{\partial \left( \porosity_\text{m} \density_{\fluidPhaseIdx} \saturation_{\fluidPhaseIdx, \text{m}} \right)}{\partial t}
    + \div \left( \density_{\fluidPhaseIdx} \darcyVelocity_{\fluidPhaseIdx, \text{m}} \right)
    &= 0,
    &&\mathrm{in} \, \domain_\text{m}, \label{eq:ex_frac_massBalanceMatrix} \\
    \frac{\partial \left( \aperture \, \porosity_\text{f} \density_{\fluidPhaseIdx} \saturation_{\fluidPhaseIdx, \text{f}} \right)}{\partial t}
    + \div \left( \aperture \, \density_{\fluidPhaseIdx} \darcyVelocity_{\fluidPhaseIdx, \text{f}} \right)
    &= \llbracket \density_{\fluidPhaseIdx} \darcyVelocity_{\fluidPhaseIdx, \text{m}} \scal \n \rrbracket,
    &&\mathrm{in} \, \domain_\text{f}, \label{eq:ex_frac_massBalanceFracture} \\
    - \frac{\relPerm{\fluidPhaseIdx,f}}{\dynVisc_{\fluidPhaseIdx}} \n^T \permeability_\text{m}^\perp
    \left( \frac{\pressure_{\fluidPhaseIdx, \text{m}} - \pressure_{\fluidPhaseIdx, \text{f}}}{\aperture/2} \, \n - \density_{\fluidPhaseIdx} \gravity \right)
    &= \darcyVelocity_{\fluidPhaseIdx, \text{m}} \scal \n, &&\mathrm{on} \, \gamma_\text{B}, \label{eq:ex_frac_ifCondition_blocking} \\
    \pressure_{\fluidPhaseIdx, \text{m}} &= \pressure_{\fluidPhaseIdx, \text{f}},
    &&\mathrm{on} \, \gamma_\text{C}, \label{eq:ex_frac_ifCondition_conductive}
  \end{align}
\end{subequations}

for $i \in \{\text{m}, \text{f} \}$ and $\fluidPhaseIdx \in \{ \water, \nitrogen \}$. Furthermore, $\density_\fluidPhaseIdx$ and $\dynVisc_\fluidPhaseIdx$ are the density and viscosity of a fluid phase $\fluidPhaseIdx$, while $\pressure_{\fluidPhaseIdx, i}$, $\saturation_{\fluidPhaseIdx, i}$ and $\relPerm{\fluidPhaseIdx, i}$ denote the fluid phase pressure, saturation and relative permeability as prevailing in subdomain $i$.
Correspondingly, $\permeability_i$ and $\porosity_i$ denote the intrinsic permeability and the porosity in a subdomain, and $\darcyVelocity_{\fluidPhaseIdx, i}$ is the respective Darcy velocity.
Finally, $\permeability_\text{f}^\perp$ denotes the normal part of the permeability in the fracture domain and $\aperture$ is the aperture of the fracture.\\
\Cref{eq:ex_frac_darcy} states that Darcy's law is valid within all subdomains, while \cref{eq:ex_frac_massBalanceMatrix,eq:ex_frac_massBalanceFracture} are the mass balance equations for the fluid phases in the matrix and fracture subdomain, respectively.
Note that the term $\llbracket \density_{\fluidPhaseIdx} \darcyVelocity_{\fluidPhaseIdx, \text{m}} \scal \n \rrbracket$ describes the jump in the normal flux of a phase $\fluidPhaseIdx$ across the fracture and acts as an additional source or sink term, caused by the interaction with the surrounding matrix.
On blocking fractures, flux and pressure continuity are enforced by means of \cref{eq:ex_frac_ifCondition_blocking}, while on conductive fractures, the pressure continuity condition \cref{eq:ex_frac_ifCondition_conductive} is used, which assumes the pressure jump across the fracture to be negligible.
The system of equations is completed by the following set of boundary conditions:

\begin{subequations}
  \label{eq:ex_frac_bcs}
  \begin{align}
    \aperture \, \density_{\fluidPhaseIdx} \darcyVelocity_{\fluidPhaseIdx, \text{f}} \scal \n
    &= 0, &&\mathrm{on} \, \Gamma_\text{f} \label{eq:ex_frac_neumann_frac}\\
    \density_{\fluidPhaseIdx} \darcyVelocity_{\fluidPhaseIdx, \text{m}} \scal \n
    &= 0, &&\mathrm{on} \, \Gamma_\text{m}^\text{N} \label{eq:ex_frac_neumann_matrix}\\
    \pressure_{\water, \text{m}} &= \pressure_\mathrm{hs}, &&\mathrm{on} \, \Gamma_\text{m}^\text{D},
                                    \label{eq:ex_frac_Dirichlet_matrix_p}\\
    \saturation_{\nitrogen, \text{m}} &= \saturation^\text{D}_\nitrogen, &&\mathrm{on} \, \Gamma_\text{m}^\text{D}.
                                    \label{eq:ex_frac_Dirichlet_matrix_S}
  \end{align}
\end{subequations}

Here, the subscript $\mathrm{hs}$ refers to hydrostatic conditions and the boundary saturation $\saturation^\text{D}_\nitrogen$ is defined as depicted in~\cref{fig:fractureExampleDomain}.
In both domains, the water pressure $\pressure_\water$ and the nitrogen saturation $\saturation_\nitrogen$ are used as primary variables and the system is closed via the constitutive relationships $\saturation_{\water, i} + \saturation_{\nitrogen, i} = 1$ and $\pressure_{\nitrogen, i} = \pressure_{\water, i} + p_{\text{c}, i}$, where $p_{\text{c}, i} = p_{\text{c}, i} ( \saturation_{\water, i} )$ denotes the capillary pressure as a function of the water saturation.
For this example, we use the van Genuchten-Mualem model for capillary pressure and relative permeability~\citep{VanGenuchten1980,mualem1976,luckner1989consistent}.
An overview over the material parameter distributions used in the subdomains is given in~\cref{tab:ex_frac_params}.

\begin{table}[ht]
  \centering
  \begin{tabular}{l l l l l l}
  \toprule
    symbol            & name               & $\Omega_\text{m}$   & $\Omega_\text{f}^\text{B}$ & $\Omega_\text{f}^\text{C}$ & unit \\ \midrule
    $\porosity$       & porosity           & \SI{0.15}{}  & \SI{0.15}{}  & \SI{0.85}{}  &  -  \\
    $k_\text{h}$             & hor.\ permeability & \SI{1e-12}{} & \SI{1e-16}{} & \SI{1e-9}{}  & \si{\square\meter} \\
    $\anisotropy$     & anisotropy ratio   & \SI{0.15}{}  & \SI{1}{}     & \SI{1}{}     & \si{\square\meter} \\
    $\permangle$      & permeability angle & \SI{-25}{}   & \SI{0}{}     & \SI{0}{}     & \si{\degree} \\
    $\alpha$          & VG parameter       & \SI{1e-3}{}  & \SI{1e-2}{}  & \SI{1e-4}{}  & \si{\per\pascal}\\
    $n$               & VG parameter       & \SI{3}{}     & \SI{2}{}     & \SI{23}{}    & - \\
  \bottomrule
  \end{tabular}
  \caption{Parameter choices for the two-phase flow through fractured porous media example. VG: van Genuchten model~\citep{VanGenuchten1980}.}
  \label{tab:ex_frac_params}
\end{table}

\begin{figure}[ht]
  \centering
  \includegraphics[width=0.99\textwidth]{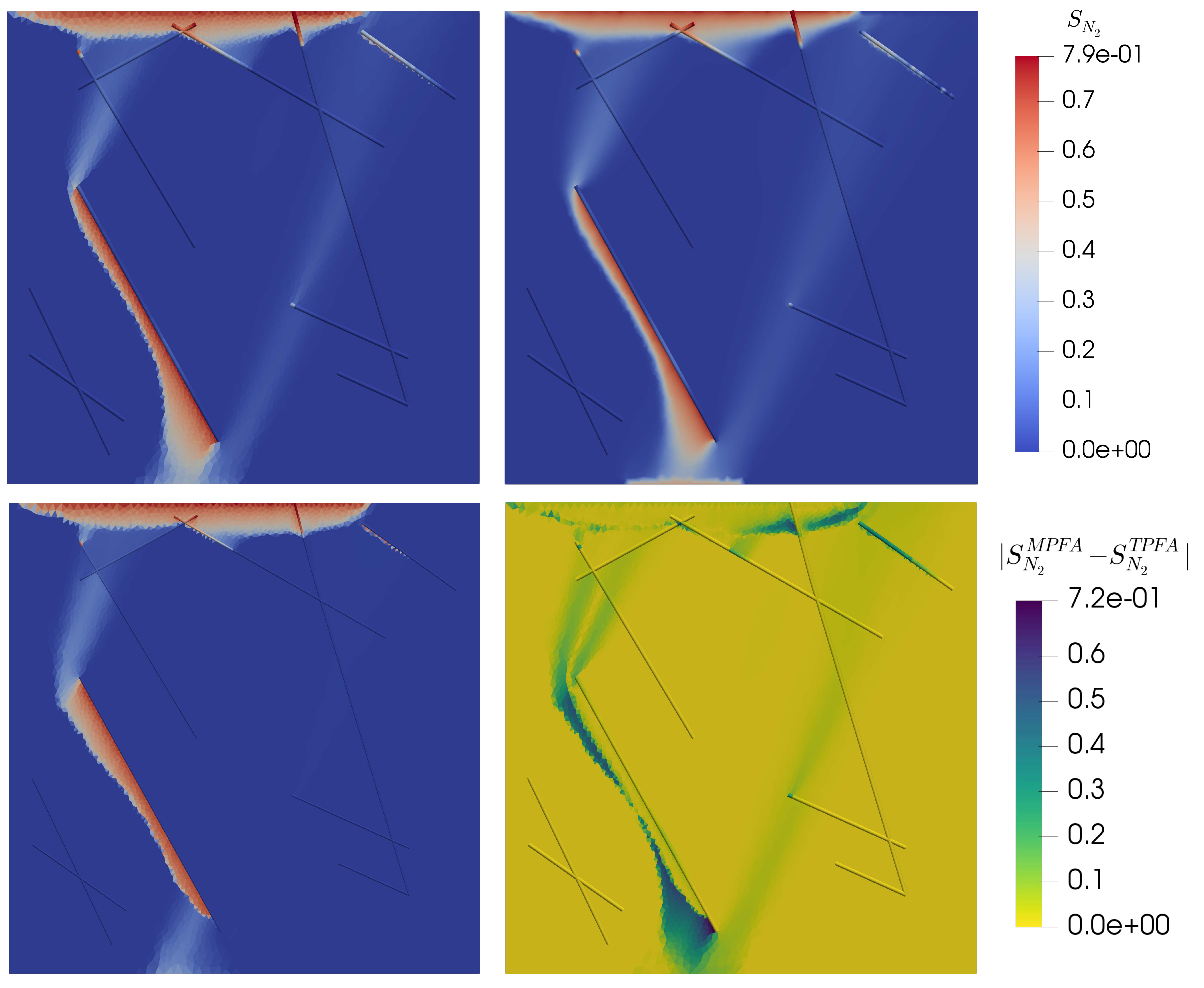}
  \caption{Nitrogen saturation distribution at the final simulation time $t = \SI{75000}{\second}$ obtained with the \textsc{mpfa-o} scheme (upper left), the \textsc{box} scheme (upper right) and the \textsc{tpfa} scheme (lower left) for the example application of two-phase flow through a fractured porous medium. The lower right image shows the difference in the saturations obtained with the \textsc{mpfa-o} and the \textsc{tpfa} scheme.}
  \label{fig:fracExampleResult}
\end{figure}

Note that the permeability of a subdomain is defined as
\begin{equation}
\permeability_i = \mathbf{R} \left( \permangle \right)^{-1} \left( \begin{matrix} k_\text{h} & 0 \\ 0 & k_\text{h}/\anisotropy \end{matrix} \right) \mathbf{R}\left( \permangle \right),
\end{equation}
where $\mathbf{R}$ is the two-dimensional rotation matrix.
In addition to that, the aperture is set to $\aperture = \SI{0.05}{\meter}$ for all fractures. The relationships of the water density and viscosity on temperature and pressure are taken from
\cite{Wagner2008Iapws}, while the nitrogen gas viscosity is modeled after \cite{reid1987}. The density of gaseous nitrogen is described with the ideal gas law.
The problem has been solved using the \textsc{tpfa}, \textsc{mpfa-o} and the \textsc{box} scheme. Switching between the different schemes is realized by changing at most two lines in the code. The grid used in this example consists of \SI{20910}{} $2$-dimensional and \SI{421}{} $1$-dimensional elements.
\Cref{fig:fracExampleResult} depicts the results for the \textsc{box} and the \textsc{mpfa-o} scheme, as well as the difference in the solution between \textsc{tpfa} and \textsc{mpfa-o} at the
final simulation time of \SI{75000}{\second}.
It can be seen that as nitrogen migrates upwards driven by buoyancy, it avoids entering the blocking fractures, which act as both capillary and hydraulic barriers,
and preferentially flows around them.
On the other hand, after entering conductive fractures, the nitrogen rapidly flows vertically and accumulates at the fracture tips due to capillary forces and the limited transport capacity of the surrounding medium.
Furthermore, the nitrogen gas accumulates below the upper no-flow boundary. The differences between \textsc{tpfa} and \textsc{mpfa-o} highlight the importance of using consistent schemes
on unstructured grids and in presence of full tensor permeabilities.
This example shows that the model is able to qualitatively capture the hydraulic effects of fractures acting as both hydraulic and capillary barriers as well as conduits.
For a more quantitative assessment of the performance of this model we refer to \citep{glaser2017discrete, glaser2019hybrid}.

\subsection{Root-soil interaction}
\label{sec:ex_root}

In the last example, we simulate root-water uptake including tracer transport.
The employed model concepts and numerical methods are described in detail in~\citep{Koch2018a}, where a similar example is discussed.
A comparable but slightly simpler example is included in the \dumux test suite\footnote{The example can be found at
\href{https://git.iws.uni-stuttgart.de/dumux-repositories/dumux/tree/master/test/multidomain/embedded/1d3d/1p2c_richards2c}{\url{git.iws.uni-stuttgart.de/dumux-repositories/dumux/tree/master/test/multidomain/embedded/1d3d/1p2c_richards2c}}}.

Root-water uptake is modeled in a cylindrical soil domain $\Omega$ with a radius of~\SI{5}{\cm} and a height of~\SI{10}{\cm}
containing a 2-week-old white lupin root system (with center-line skeleton $\Lambda$),
reconstructed from MRI data~\citep{Schrder2014}. The soil domain $\Omega$ is discretized with an unstructured tetrahedral grid refined around the root system, using \dunemodule{dune-uggrid}, while
$\Lambda$ is represented by an independent grid of line segments forming a root network, using \dunemodule{dune-foamgrid}~\citep{foamgrid}.
At the root collar a transpiration rate
of $r_\text{T} = \operatorname{min}\{\SI{2.15e-8}{\kg\per\s}, r_{\text{T}, \text{c}}\}$ is prescribed, where $r_{\text{T}, \text{c}}$ is the transpiration rate for which
the root collar pressure is $p_{\text{r}, \text{c}} = \SI{-1.4}{\mega\Pa}$ (wilting point pressure).
The soil domain contains a tracer of initially uniformly distributed concentration ($x^\kappa_\text{w} = \SI{3e-7}{}$). The tracer (binary diffusion coefficient in water $D^\kappa_\text{w} = \SI{2.3e-9}{\m\per\square\s}$) does not enter the roots. As a consequence, it
is expected to accumulate at the locations of the largest root water uptake.
The soil flow is modeled with the Richards equation, and the flow inside the root xylem is modeled using a bundle-of-tubes approach, resulting in a Darcy-type flow model, cf.~\citep{Koch2018a}.
The nonlinear PDE system
\begin{subequations}
\label{eq:ex_rootsoileq}
\begin{align}
\frac{\partial \left( \phi_\text{s} S_\text{w} \rho_{\text{m}, \text{w}} \right) }{\partial t} -\div \left(\rho_{\text{m}, \text{w}}\boldsymbol{v}_\text{s}
\right) &= q_\text{w} \delta_\Lambda &\text{in }\Omega, \label{eq:ex_soil}\\
\frac{\partial \left(\phi_\text{s} S_\text{w} \rho_{\text{m}, \text{w}} x^{\kappa}_\text{w} \right)}{\partial t}
-\div \left(x^{\kappa}_\text{w} \rho_{\text{m}, \text{w}} \boldsymbol{v}_\text{s}
+ \rho_{\text{m}, \text{w}} D_{\text{eff}}^{\kappa} \grad x^{\kappa}_\text{w}
\right) &= 0 &\text{in }\Omega, \label{eq:ex_soiltracer}\\
\frac{\partial \left(\phi_\text{r}\rho_{\text{m}, \text{w}} A_\text{r} \right)}{\partial t}-
\frac{\partial}{\partial \zeta} \left(\rho_{\text{m}, \text{w}}v_\text{r}
\right) &= -q_\text{w} &\text{in }\Lambda, \label{eq:ex_root}
\end{align}
\end{subequations}
where
\begin{align*}
\boldsymbol{v}_\text{s} &= - \frac{k_{\text{rw}}}{\mu_\text{w}} K \left(\grad p_\text{w} + \rho_\text{w} g \grad z \right),\\
v_\text{r} &= - K_{\text{ax}} \left(\frac{\partial p_\text{w}}{\partial \zeta} + \rho_\text{w} g \frac{\partial z}{\partial \zeta} \right),\\
q_\text{w} &= -2 \pi R k_\text{rw} K_\text{rad} \left( \bar{p}_\text{s} - p_\text{r} \right)\rho_{\text{m}, \text{w}},
\end{align*}
with the average soil pressure $\bar{p}_\text{s} = \frac{1}{2\pi} \int_0^{2\pi} \! p_\text{s}\vert_R \text{d}\theta$,
and an analogous definition for $\bar{x}^{\kappa}_{\text{w}, \text{s}}$~\citep{Koch2018a,dangelo2008},
is solved for water pressure $p_\text{w}$ and tracer mole fraction $x^{\kappa}_\text{w}$ for a period of \SI{3}{\day} with a maximum time step size of \SI{1}{\hour}.
The symbols and parameter values are given in~\cref{tab:rootsoilparams}.
The subscripts r and s denote root and soil quantities wherever they
need to be distinguished.
\Cref{eq:ex_soil,eq:ex_soiltracer} are spatially discretized using the \textsc{box} scheme, while~\cref{eq:ex_root} is discretized with the \textsc{tpfa} scheme. The corresponding coupling manager is generic, in the sense that it can compute the coupling stencils between the root and the soil domain if the discretization schemes in both domains differ. The \textsc{box} scheme is superior to the \textsc{tpfa} scheme in the soil domain for the presented case, because it usually uses less degrees of freedom on tetrahedral grids, and it is well-known that the \textsc{tpfa} scheme is not consistent on non-K-orthogonal grids~\citep{Schneider2018}.
In comparison with~\cref{code:intersection}, the procedure to compute the coupling stencil is slightly more complex for this example, since the integration of the source term $q_\text{w}$
requires the evaluation of quantities on the root segment surface, resulting in non-local stencils. For details on the discretization method for this mixed-dimension embedded problem, we refer to~\citep{Koch2018a}.
The required classes and functions can be found in the \dumux repository's \texttt{dumux/multidomain/embedded} folder.

\begin{figure}[ht]
  \centering
  \includegraphics[width=\textwidth]{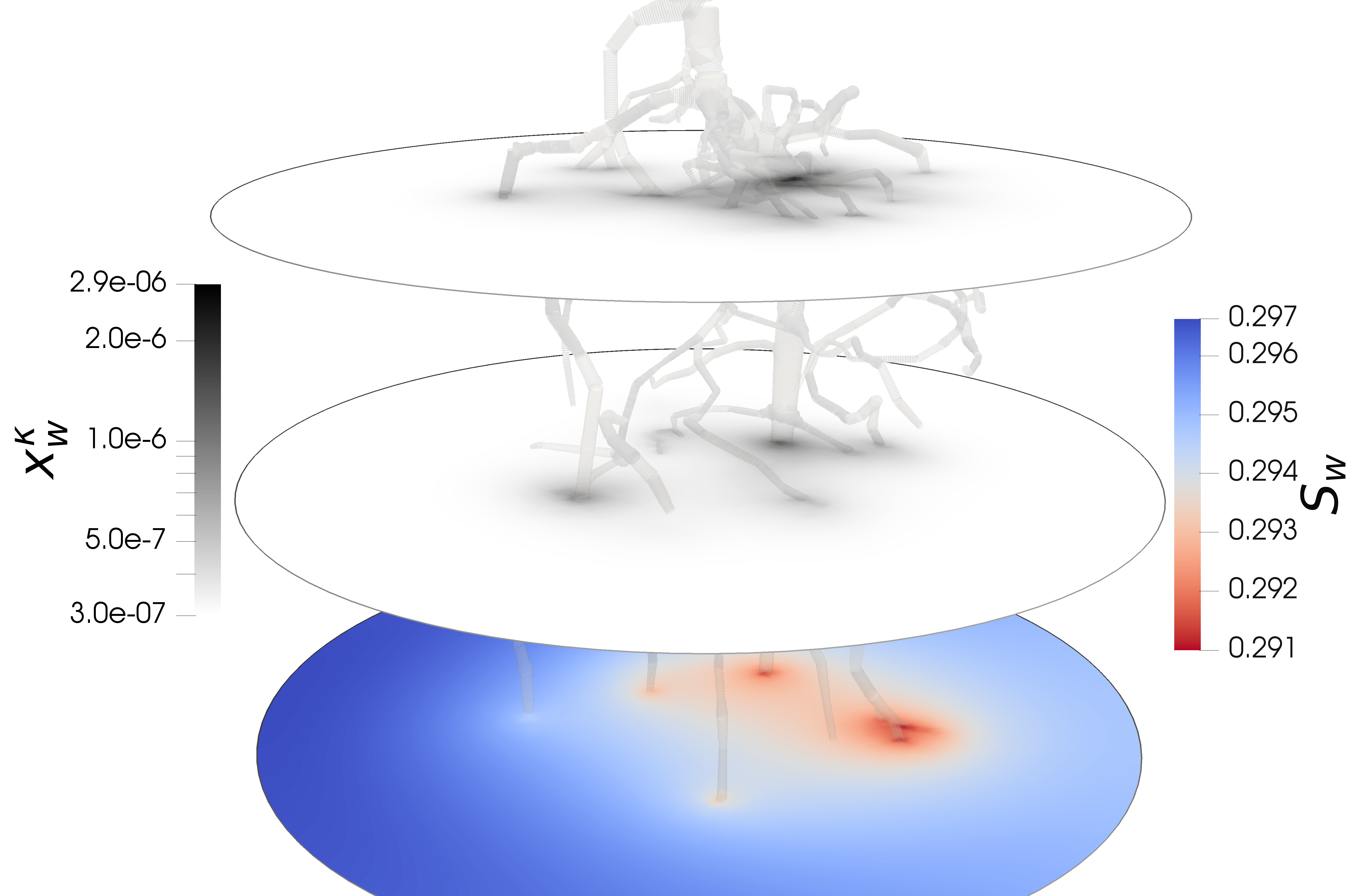}
  \caption{Simulation of root-water uptake of a white lupin and simultaneous tracer transport in the soil, shown at $t=\SI{3}{\day}$. Three horizontal
           cuts through the soil domain are shown.
           The tracer, with mole fraction $x^\kappa_\text{w}$, is not taken up by the roots and accumulates, particularly where the root water uptake rate is highest.
            On the bottom slice, the water saturation $S_\text{w}$ is shown. The saturation slightly decreases close to the roots. Its spatial gradient
            depends on the flow resistances in soil and root, the current water distribution, and the prescribed transpiration rate $r_\text{T}$. }
  \label{fig:rootsoil}
\end{figure}

The resulting spatial distribution of water and the tracer is shown
in~\cref{fig:rootsoil}. A tracer accumulation in close vicinity
to the roots is evident. This type of simulation may present a valuable
tool to interpret experimental tracer data. Due to the non-linear
relation of the tracer concentration distribution to the water uptake rate and its dependence on water distribution, it is otherwise difficult to obtain quantitative result on local root water uptake rates from tracer concentration measurements.

\begin{table}[ht]
  \centering
  \begin{tabular}{l l l l}
  \toprule
    symbol & name              & value  & unit \\ \midrule
    $\phi_\text{s}$ & soil porosity                 & \SI{0.4}{} & - \\
    $\rho_{\text{m}, \text{w}}$ & molar density of water & \SI{5.55e4}{} & \si{\mol\per\cubic\meter}  \\
    $D_{\text{eff}}^{\kappa}$ & effective diffusion coefficient & \citep[][Eq. (16)]{Koch2018a} \\
    $\phi_\text{r}$ & root porosity                 & \SI{0.4}{} & - \\
    $A_\text{r}$ & root cross-sectional area                 &  & \si{\square\meter} \\
    $\zeta$ & local axial root coordinate   &  & - \\
    $S_\text{w}$ & water saturation                 & $S_\text{w}(p_\text{w})$, \citep{VanGenuchten1980}~\citep[][Eq. (8)]{Koch2018a} & - \\
    $k_{\text{rw}}$ & relative permeability  & $k_{\text{rw}}$ \citep{VanGenuchten1980,mualem1976}~\citep[][Eq. (9)]{Koch2018a} & -  \\
    $\alpha$ & van Genuchten parameter & \SI{2.956e-4}{} & \si{\per\Pa} \\
    $n$ & van Genuchten parameter & \SI{2.0}{} & - \\
    $S_\text{wr}$ & residual water saturation & \SI{0.1}{} & - \\
    $g$ & gravitational acceleration & \SI{9.81}{} & \si{\m\per\square\s} \\
    $K_\text{ax}$ & axial root conductivity & \SI{5.10e-17}{} & \si{\m^4\per\Pa\per\s} \\
    $K_\text{rad}$ & radial root conductivity & \SI{2.04e-11}{} & \si{\m\per\Pa\per\s} \\
  \bottomrule
  \end{tabular}
  \caption{Symbols and parameter values for the root-soil interaction example.}
  \label{tab:rootsoilparams}
\end{table}

\section{Current limitations and perspectives}
\label{sec:future}

As a research code under active development, \dumux currently
has certain limitations, from which many may be resolved in future versions.
We briefly mention some of the limitations.

Multi-domain simulations in \dumux currently do not run in
parallel when involving multiple grids. In \dune, the \dune grid
implementations are responsible for managing MPI-based distributed memory parallelism.
When having two or more grid instances, data has to be communicated between grids.
First steps towards such a feature within the \dune framework
have been undertaken with the \dunemodule{dune-grid-glue} module~\citep{Bastian2010}.
Efficient load balancing in complicated multi-domain, or mixed-dimensional
setups is challenging and is an active field of research targeted by codes such as
the MOOSE framework~\citep{Gaston2009}, or preCICE~\citep{preCICE}.

\dumux currently supports forward and backward Euler time discretizations only.
Implementing other time discretization schemes requires some non-trivial refactoring
of the assembly process.

As discussed in~\cref{sec:design}, \dumux uses a C++ programming technique based
traits of tags to define what we call \textit{properties} of a model. Due to the
dependency on a tag, many of such properties (often corresponding to C++ types) are grouped together. Some classes
have such a tag as a template argument (see \cpp{FVAssembler} in~\cref{code:main}, l.~44).
While this significantly reduces the number of template arguments of a class,
it leads to an nontransparent way of injecting dependencies. Furthermore,
\cpp{FVAssembler<Tag1>} and \cpp{FVAssembler<Tag2>} are different types, even if all
type traits (properties) extracted from the tags are identical. Modularity and re-usability
of such classes is impeded. Thus, the use of tags as template arguments has been significantly
reduced in \dumux 3 in comparison with previous versions. A medium-term goal is to replace
all such classes with classes with explicit dependencies. The property technique is also
the main reason, why it is currently difficult to add useful Python wrappers
(a feature desired by many \dumux users) for \dumux.

The focus of \dumux is on providing many usable and extensible models, and less
on flexible linear algebra. While efficient and flexible data structures
and easy-to-use and extensible, preconditioned
linear solvers are available through \dunemodule{dune-istl}~\citep{duneistl}, \dumux
currently does not facilitate algebraic manipulations of linear equation systems,
such as different orderings of the Jacobian matrix, quasi-Newton schemes, or complex linear solver strategies
based on matrix decomposition.

The future development of \dumux will continue to reflect the most recent advances in the field of porous medium research within the \dumux community.
For example, the goal of a current project is the improvement of solvers for free-flow and shallow water models coupled with porous medium flow models.
Pore-network models for single- and multi-phase flow will be included into the \dumux framework in an upcoming release, including static and dynamic approaches.
In collaboration with the development team of preCICE~\citep{preCICE}, the model coupling capabilities of \dumux with internal and external modules will be
further improved.
The usability and archivability of \dumux will be further improved within the aforementioned project SusI which constitutes a step towards reproducible research.

\section*{Acknowledgements}

This work was financially supported by the German Research Foundation (DFG), within the Cluster of Excellence in Simulation Technology (EXC 310),
the Collaborative Research Center on Interface-Driven Multi-Field Processes in Porous Media (SFB 1313, Project No. 327154368), and
the DFG project Sustainable infrastructure for the improved usability
and archivability of research software on the example of the porous-media-simulator \dumux (Project No. 391049448).
Furthermore, we would like to acknowledge all individuals that have made and will make contributions to the open-source
project \dumux in any way. The successful development is only possible due to input from the \dumux user community.

\section*{Author roles}
T. Koch, D. Gläser, K. Weishaupt wrote and proof-read this manuscript, conceptualized, implemented and visualized the presented examples,
and curated the data and software. Furthermore they are the authors with the most code contributions (measured in Git commits to the \dumux repository since the previous release 2.12).
B. Flemisch wrote and proof-read parts of this manuscript, continuously contributes code, is the administrative head and coordinator of the \dumux project, and acquired funding.
All other authors (listed in alphabetical order) committed source code to the \dumux repository since release 2.12 and
thus contributed to the development and increasing quality and number of features of the \dumux research software framework, version~3.

\bibliography{dumux}
\bibliographystyle{elsarticle-num}

\end{document}